\newcommand{\bea}{\begin{eqnarray}}
\newcommand{\eea}{\end{eqnarray}}
\def \rt {{\rm t}}
\def \ci{\cite}

\newcommand{\p}[1]{(\ref{#1})}
\newcommand{\bt}[1]{{\bar t}}

\def\dg{\dagger}

\def \sm {{sigma model\ }}

\newcommand{\commentout}[1]{}

\documentclass[12pt]{article}

\usepackage{amsmath}
   \usepackage{amscd}
   \usepackage{amsfonts}
   \usepackage{amssymb}
   \usepackage{amsthm}
   \usepackage{epsfig}

\def\be{\begin{equation}}
\def\ee{\end{equation}}
\def\ba{\begin{eqnarray}}
\def\ea{\end{eqnarray}}

\def\dg{\dagger}
\def\a{\alpha}
\def\b{\beta}
\def\e{\varepsilon}
\def\p{\partial}

\def\R{{\bf R}}

\def\Tr{{\rm  Tr}}

\def\ket[#1]{\left|#1\right>}
\def\bra[#1]{\left<#1\right|}

\numberwithin{equation}{section} 
\parskip=\medskipamount

\input epsf
\usepackage{epsfig,color,latexsym}
\def \del{\partial}
\def \a {\alpha}

\def\g{\gamma}
\def\s{\sigma}

\def\ov{\over}
\def\la{\label}

\def\J{{\cal J}}

\def\b{\beta}
\def\l{\lambda}

\def \adss{$AdS_5 \times S^5$\ } 

\def \r { \rho}
\def \sql {\sqrt{\lambda} }

\def \vp {\varphi}

\def \ov {\over}
\def \s{\sigma}

\def \ha {{1 \over 2}}

\def \la{\label}

\def \k {\kappa}
\def\foot{\footnote}

\def \J {{\cal J}}
\def \L {\Lambda}

\newcommand{\rf}[1]{(\ref{#1})}
\renewcommand{\theequation}{\thesection.\arabic{equation}}
\renewcommand{\thefootnote}{\fnsymbol{footnote}}

\def\appendix#1{
  \addtocounter{section}{1}
  \setcounter{equation}{0}
  \renewcommand{\thesection}{\Alph{section}}
  \section*{Appendix \thesection\protect\indent \parbox[t]{11.15cm}
  {#1} }
  \addcontentsline{toc}{section}{Appendix \thesection\ \ \ #1}
  }

\def \ci {\cite}

\def \foot {\footnote}
\def \bi{\bibitem}

\def \ha {{1 \over 2}}
\def \td {\tilde}

\def \ci{\cite}

\def \rH {{\rm H}} \def \rE {{\rm E}}

\def \tl {{\tilde \lambda}}
\def \sql{\sqrt{\l}}
\def \inti {{\int^{2\pi}_0 {d \sigma \ov 2 \pi}}}
\def \x { y}
\def \be {\bea}
\def \ee {\eea}
\def \d {\partial}
\def \K {{\rm S}}
\def \el {\ell}
\def \Tr {{\rm Tr}}

\begin{document}
\thispagestyle{empty}
\def\thefootnote{\fnsymbol{footnote}}\begin{flushright} 
hep-th/0404133\\ 
Imperial/TP/3-04/10
\end{flushright}\vskip 0.5cm\begin{center}
\LARGE{\bf  Large spin limits of AdS/CFT 
 and  generalized    Landau-Lifshitz equations}
\end{center}\vskip 0.8cm\begin{center}{\large B. Stefa\'nski, jr.$^{1,}
$\footnote{E-mail address: {\tt b.stefanski@ic.ac.uk}}
and A.A. Tseytlin$^{2,1,}$\footnote{Also at Lebedev Institute, Moscow. }
}\vskip 0.2cm{\it $^1$ Theoretical Physics Group,  Blackett  Laboratory, \\
Imperial College,\\ London SW7 2BZ, U.K.}
\vskip 0.2cm
{\it $^2$  Smith Laboratory, The Ohio State University, \\
Columbus, OH 43210-1106, USA}
\end{center}
\vskip 1.0cm
\begin{abstract}\noindent 
We consider  $AdS_5\times S^5$  string states with  several 
large
angular momenta along $AdS_5$ and  $S^5$ directions
which are dual to single-trace Super-Yang-Mills (SYM) operators  built out of 
chiral 
combinations of scalars  and covariant derivatives.  
In particular, we focus on the $SU(3)$ sector  (with three spins in $S^5$) and
 the $SL(2)$ sector 
(with one spin in $AdS_5$ and one in $S^5$), generalizing  recent work 
hep-th/0311203 and  
hep-th/0403120 on the $SU(2)$ sector with two spins in $S^5$.
We show that,  in the large spin limit and at leading order in 
the effective coupling expansion,   the string 
sigma model equations of motion
reduce to matrix Landau-Lifshitz equations.
 We then demonstrate  that
  the coherent-state expectation value of the one-loop  SYM 
 dilatation operator 
restricted to  the corresponding sector of single trace operators is 
also effectively described by the same equations. This implies a universal leading
order equivalence between string energies and SYM anomalous dimensions,
as well as a matching of integrable structures.
We also discuss the more general 
5-spin sector  and comment  on $SO(6)$  states 
dual to non-chiral scalar  operators. 
\end{abstract}

\vfill

\setcounter{footnote}{0}
\def\thefootnote{\arabic{footnote}}
\newpage

\renewcommand{\theequation}{\thesection.\arabic{equation}}
\section{Introduction}
Following  earlier suggestions~\ci{bmn,gkp} to study sub-sectors 
of string states with large quantum numbers, it was proposed in~\ci{ft2,ft3} 
(see also~\ci{tse2} for a review) 
that spinning string states  with 2+3 angular momenta $(S_1,S_2; J_1,J_2,J_3)$
in \adss should be dual to local operators\foot{Here and in similar expressions 
below 
 $+\dots$ stands  for  all other orderings of the fields and derivatives 
  inside the trace.} 
\be
O_{J,S}=\mbox{\rm Tr}(D_{1+i2}^{S_1} D_{3+i4}^{S_2}
\Phi_1^{J_1} \Phi_2^{J_2} \Phi_3^{J_3}) + \dots
\ee
in the large $N$ limit of the maximally supersymmetric Super-Yang-Mills
 (SYM) theory on $R^4$. 
In particular, the classical string energy 
\be
E= J + S  + c_1 \frac{\l}{  J} 
+ c_2 \frac{\l^2}{  J^3}
+ ... \ , \ \ \ \ \ \ \ \ S=S_1+S_2\ , \ \ \ \ \ \ \ 
 J= J_1 + J_2 + J_3 \   , \ee
  which happens  to have a regular expansion 
in powers of $\tl = \frac{\l}{ J^2}$ at large  total $S^5$  angular momentum 
$J$  was suggested to reproduce the  anomalous dimensions
of the corresponding operators computed in the same limit
\be   \la{lim}
J \to \infty\   , \  \ \ \ \ \ \   \ \ \tl= {\l \ov J^2}={\rm fixed} \ .  \ee
 In the ``$SU(2)$'' sub-sector of  2-spin states  represented by strings 
positioned at the center of $AdS_5$ and rotating in two out of three 
planes, i.e. having $S_1,S_2=0, J_3=0, \ J_1,J_2\not=0$, 
this proposal was explicitly confirmed   in~\ci{bmsz,ft4,afrt,bfst}
by comparing the leading coefficient $c_1=c_1({J_1\ov J_2})$  
in the string energy for particular 2-spin string solutions 
with  the  one-loop anomalous dimensions of  the 
scalar SYM operators  Tr$(\Phi_1^{J_1} \Phi_2^{J_2}) + ... $.
The anomalous dimensions were  
 computed~\ci{bmsz,bfst} using the $SU(2)$  
  Heisenberg  XXX$_\ha $ spin chain interpretation of 
the one-loop anomalous dimension
 in the corresponding 
scalar sector~\ci{mz1} and taking a  thermodynamic $J \to
 \infty$  limit of the Bethe ansatz solution for the eigenvalues
($1 \over J$ corrections should correspond to quantum string \sm 
corrections).
This ``one-loop'' 
agreement was   extended to the ``two-loop'' ($c_2$-coefficient) 
level~\ci{serb}
and also demonstrated (at one and two loop orders) 
for  all classical solutions~\ci{kmmz}
 using integrable model/Bethe ansatz techniques. 
A similar approach was applied also 
 to theories with less supersymmetry~\cite{lesssusy}
 and to examples   with
both open and closed strings~\cite{openstring}.

A more universal and potentially deeper  understanding of  how this 
relation between (classical)  string  theory and 
(quantum)  gauge theory arises in the $(J_1,J_2)$ sector 
was recently presented in~\ci{kru,krt}. In this approach one identifies
 a collective coordinate $\a$ associated to $J$ 
and eliminates it from the dynamics. As a result,   
 the action of  the  classical  bosonic \adss 
\sm can be rewritten in the limit $J \to \infty,\ \tl  <1$ 
as a non-relativistic two-dimensional theory for the ``transverse'' string 
 coordinates $n_i$ \ ($n^2=1$),  with the structure 
\be \la{lags}
L=  J \bigg( C_i(n) \d_0 n_i  - \bigg[\tl a_0 (\del_1 n)^2  + 
\tl^2  [ a_1 (\del^2_1 n)^2  + a_2 (\del^2_1 n)^4 ] 
+ ...\bigg]\bigg) \ , \ee 
where $C_0\equiv C_i(n) \d_0 n_i $ may be interpreted as  a WZ-type term 
($C_i$ is a monopole potential, $dC_i=- \ha \epsilon_{ijk} n_j dn_j$).
It was shown that  this action agrees precisely  
(at order $\tl$ \ci{kru} and $\tl^2$ \ci{krt}) 
with the corresponding low-energy effective action of the $SU(2)$ 
ferromagnetic spin chain with the Hamiltonian $H$ given by 
the sum of the one-loop \ci{mz1} and two-loop \ci{bks}
 dilatation operators. The leading term in the latter  action 
is determined \ci{fra,ran} by the coherent state \ci{pere} 
($ \bra[n] \s_i \ket[n]=
n_i$) 
expectation value of $H$.
The agreement at the level of two-dimensional actions implies a matching
between energies of all string/spin chain solutions 
and gives a  direct relation between integrable 
structures (observed earlier  using Bethe ansatz  approach 
in \ci{as,emz,kmmz}).

It is of obvious interest to extend the approach of~\ci{kru,krt}
to other sectors of rotating string states.
 To do this one has to identify  subsectors of operators of 
  the gauge theory which are closed under renormalization 
  at least at one-loop. Here we will be  interested only in the 
  bosonic subsectors.
   Apart from the $SU(2)$ sector which is closed to
    all loop orders, other such sectors 
    are\footnote{We are grateful to Niklas Beisert for 
    a discussion and clarification of this issue.}

(i) the three-spin 
``$SU(3)$'' sector 
 of string configurations  with  all 
three $S^5$ angular momenta 
$(J_1,J_2,J_3)$ being non-zero. These are 
dual to more general chiral operators 
Tr$(\Phi_1^{J_1} \Phi_2^{J_2} \Phi_3^{J_3}) + ... $
(which form a set closed only under one-loop renormalization~\ci{bks})

(ii) the two-spin  ``$SL(2)$'' sector of string
configurations  with one $AdS_5$ spin ($S=S_1$) and one $S^5$ 
angular momentum ($J=J_3$), which are dual to operators 
Tr$(D_{1+i2}^S \Phi^J) +...$ (forming a set  closed 
under renormalization to all orders~\ci{bei,bs,beg}).

(iii) the three-spin ``$SU(1,2)$'' sector of string
 configurations with two $AdS_5$ angular momenta 
 $(S_1,S_2)$ and one $S^5$ 
spin ($J=J_3$), which are dual to operators 
Tr$(D_{1+i2}^{S_1}D_{3+i4}^{S_2} \Phi^J) +...$ 
(which form  a set closed under one-loop renormalization).

\noindent Operators in other sectors carrying more general
 configurations of non-zero spins $(S_1,S_2,J_1,J_2,J_3)$ 
 mix with fermionic operators already  at one loop 
 and would require to consider superspin  chains \ci{bs,bb}
 and to include fermions on the 
 string sigma-model side.

Earlier  results  demonstrating  the 
matching of  the leading $c_1$ coefficient in the string energy 
for certain string solutions 
and the corresponding 
one-loop anomalous dimensions were  already  found
in the  $SU(3)$ case in \ci{emz,kri} 
(see also \ci{engquist})  and in the   $SL(2)$ case 
in \ci{bfst}.  

In the present paper we show that for the $SU(3)$ and $SL(2)$ sectors, 
the  one-loop equivalence holds universally   at the level 
of the corresponding effective two-dimensional actions. 
This implies a manifest  agreement of the leading-order coefficients 
in the classical string energy and in the one-loop SYM 
anomalous dimensions for all possible configurations 
with given charges and also guarantees a matching of  other conserved
charges (i.e. the equivalence of integrable structures).
 The equations that follow from the resulting leading-order
2-d  action  are  matrix generalizations  of the 
Landau-Lifshitz equations. 

On the string side (section  2), we use a Hopf-type 
parametrization of $AdS_5$ and $S^5$ metrics separating a single common  phase 
direction. On the SYM side, 
in the $SU(3)$ sector (section 3)  our starting point will be  the general 
expression for the one-loop  dilatation operator in the 
scalar sector as a Hamiltonian of an $SO(6)$ spin chain~\ci{mz1} 
which we restrict to the 
chiral $SU(3)$ states  and  compute its expectation value in 
the $SU(3)$ coherent state formalism.
In the $SL(2)$ sector (section 4) 
 we use the expression for the one-loop 
dilatation operator as a Hamiltonian of the XXX$_{-\ha}$ 
 $SL(2)$ spin chain derived in~\ci{bei}. Here the length of the spin chain is $J$, 
but the number of states $S$ at each site  can be arbitrarily large. 
We define the relevant coherent state and use it to find the associated semi-classical 
 action by computing the expectation value of the spin chain Hamiltonian.

Both $SU(3)$ and $SL(2)$ dilatation operators are special cases  of the  most 
general
 $PSU(2,2|4)$  one-loop dilatation operator \ci{bs,bb}.
In the general sector involving non-chiral operators  one does not expect a
direct semi-classical relation to  string theory:
string $\alpha'$ corrections are expected to be important in this 
case even in the $J \to \infty$ limit so one should be comparing 
to the  full quantum string theory. 
  Still, it may be of interest to study a \sm 
that represents a  semi-classical coherent state effective action 
of the  $PSU(2,2|4)$ spin chain.
In section 5 we comment on the coherent state
expectation value of the $SO(6)$ spin chain Hamiltonian 
representing the one-loop dilatation operator in the sector of general 
(non-chiral)  scalar operators \ci{mz1}.
Some conclusions will be  summarized in section 6.

In Appendix A we present a lightning review of coherent 
states following~\ci{pere} and discuss more explicitly the 
coherent states for $SO(6)$. Some of the computational details
 relevant to Section 4 are presented in Appendix B.

\section{From  \adss  string  sigma model to  Landau-Lifshitz equations}

In this section we give a procedure 
for ``rearranging'' the  classical action of \adss string \sm 
in the large $S^5$ spin limit which generalizes the one in 
\ci{kru,krt} from the 2-spin sector to more general 
configurations with two  $AdS_5$ spins and three  $S^5$ spins.
 It  
 leads to a non-relativistic 
2-d action for ``transverse'' string coordinates 
which is first order in time derivatives
 and  higher order  in spatial 
derivatives. The equations of motion that follow from the leading-order term in this action are the generalized  Landau-Lifshitz equations.

\subsection{Parametrization of \adss}

As in \ci{ft2},  we parametrize the   $AdS_5\times S^5$ metric
in terms of 3+3 complex coordinates 
(we assume summation over repeated indices $i,j=1,2,3$ and  ${}^*$ 
denotes  complex conjugation) 
\bea
ds^2= dY^*_i dY^i+ dX^*_i  dX_i \,,
\eea
where $Y^i=\eta^{ij}Y_j$ with $\eta^{ij}=\mbox{diag}(-1,1,1)$ and
\be 
 Y^*_iY^i=-1\,,\qquad\qquad X^*_i X_i=1\,.
\ee
Introducing new coordinates $\x,\a$, $V_i$, $U_i$ as follows
\be
Y_i=e^{i\x}V_i\,,\qquad\qquad
X_i=e^{i\alpha}U_i\,, \ \ \ \ \ \ \ \ 
  V^*_iV^i=-1\,,\qquad   U_i^* U_i=1\, ,  
\ee
the metric becomes
\ba
ds^2&=&-(d\x+B)^2+dV^*_i dV^i + B^2+(d\alpha+C)^2+dU_i^* dU_i -C^2\nonumber \\
&=&-(d\x+B)^2+  (d\alpha+C)^2 +  D^*V^*_i DV^i + D^*U_i^* DU_i  \,,
\la{met}
\ea
where
\be
B=iV^*_idV^i\,,\ \ \  DV_i=dV_i-iBV_i\,,\ \ \ \
C=-i U_i^*dU_i\,,\ \ \  DU_i=dU_i-iCU_i\,.
\ee
 Here $\x$ and $\a$ are ``overall'' phases of  $AdS_5$ and $S^5$ 
and 
$U_i$ and $V_i$ are  projective space coordinates:
the metric is invariant under a simultaneous shift of $\x$ and rotation of 
$V_a$,
as well as a shift of $\a$ and rotation of $U_i$.
The $U(1)$ connections $B$ and $C$ are real. 
This parametrization corresponds to a Hopf $U(1)$ fibration of $S^5$  over 
$CP^2$
and a similar  fibration of $AdS_5$ over a non-compact version of $CP^2$. 
Indeed, 
$ds^2_4=  DU_i^* DU_i$ is the Fubini-Studi metric on $CP^2$ 
and $K=\ha dC$ is the covariantly constant  K\"ahler form on 
$CP^2$.\foot{For other 
applications of similar  parametrizations see, e.g., 
\ci{duf,pops}.}

It is useful also to recall the relation to the standard angular parametrization 
of \adss: ($Y_i=(Y_0,Y_1,Y_2), \  X_i = (X_1,X_2,X_3)$)
\be \la{an}
Y_0 = \cosh \r \ e^{it} \ , \ \ \ \  Y_1 = \sinh \r \ \cos \theta \ e^{i \phi_1} 
\ , 
\ \ \   \  Y_2 = \sinh \r \ \sin \theta \ e^{i \phi_1} \ , \ee
\be \la{gl}
X_1  = \sin \g \ \cos \psi \  e^{i\vp_1 } \ , \ \ \ \  
X_2 = \sin \g  \ \sin \psi \ e^{i \vp_2} \ , 
\ \ \   \  
X_3 = \cos \g  \ e^{i \vp_3} \ . \ee

\subsection{Large spin limit of \adss  sigma model}

Let us now consider the string \sm action for the metric \rf{met}:
\be 
L = -\ha \bigg[ -(\del_a \x+B_a)^2+  (\del_a \alpha+C_a )^2 +  D^*_a V^*_i 
D^aV^i +
 D^*_a U_i^* D^a U_i\bigg]   \,.
\la{meta}
\ee
It is clear from \rf{an},\rf{gl}  that 
the  conserved charge corresponding to 
translations in $\x$ is the difference (or sum, depending on conventions)
 of $AdS$ energy and two $AdS$ spins, i.e.
 $E-( S_1 +S_2)$,  while the charge  corresponding to 
$\a$ is  the total $S^5$  angular momentum $J\equiv J_1 + J_2 + J_3$
(see \ci{ft2}).
Consider string configurations  for which 
$\x \approx \a$, in other words interpret $\x$ as time and boost  along $\a$. Such configurations carry large spin with
$
E= J + S + O({\l\ov J})$.
Note that we are using 
the existence  of one common ``angle'' in 
 $AdS_5$ ($\x$) and one  in $S^5$ ($\alpha$) which enter the metric\rf{met}
  with the
 {\it opposite} signs. 
 This procedure does not apply if all 
 the $S^5$ directions are trivial, i.e. if 
 the string were  moving only in
 $AdS_5$. Indeed, in this case it is known \ci{gkp,ft1,ft2}
  that the string energy 
 does not have a regular expansion in $\l$ at large spin;
  the discussion that follows applies only to string configurations 
 with   at least one angular momentum component in $S^5$ direction.

 Below  we will  be interested only in the leading
``one-loop''  term in the large spin or small $\tl={\l\ov J^2}$ expansion 
of the action. The reason is that, 
as already mentioned above,  while   the two-spin 
operators from  the $SU(2)$  sector discussed 
in \ci{kru,krt} (and also two-spin operators from $SL(2)$ sector)
  are closed under renormalization to all  loop orders, 
this is not so for more general 3-spin operators, i.e. 
  one is able to compare to (one-loop) SYM theory only the leading-order 
term in expansion of the sigma model action.

Following \ci{kru,krt}, we gauge away  the ``longitudinal'' 
coordinates $\x$ and $\a$ 
and arrive at an action  for the ``transverse'' 
coordinates $V_i$ and $U_i$
(which, in particular,  determine a ``shape'' of rigid rotating strings 
in the solutions discussed in \ci{ft2,ft4,afrt,art}). 
In the  general case of the $(S,J)$ sector of configurations 
 with all 5 spins non-zero
the choice of a useful 
  gauge fixing  procedure appears to be  non-trivial
(see equation~\eqref{twe} below).
   One possibility is to start 
with a first-order form of the action 
as in \ci{krt}
and fix the momenta  corresponding to $\x$ and $\a$ 
 to be  homogeneous. 
Since we are interested only in the leading order correction
it should be  sufficient to  use  the conformal gauge 
supplemented by a  condition like $\x=\k \tau$ or $\a=\J \tau$ 
(analogous to  $x^+=p^+ \tau$) 
that fixes the remaining conformal diffeomorphisms. 
The difficulty in choosing such 
a simple gauge for the  general case of four- or 
five-spin configurations  should be effectively related 
 to    the fact (mentioned in the Introduction) 
 that  the corresponding more general 
gauge theory  operators built out of
 chiral scalars and covariant derivatives do not form a closed 
 subsector  mix  with operators involving fermions
  already at one loop.

Instead of attempting to address the most general case of 
all $S_n$ and $J_i$ 
being non-zero
here we shall concentrate on two special three-spin cases
(related by an  analytic continuation \ci{bfst}):

(i) $ S_1,S_2=0, \ J_1,J_2,J_3\not=0$,\ \  \ \ \  i.e.  $Y_0 = e^{it}, \ 
Y_1,Y_2=0, \ y=t$

(ii) $ J_1,J_2=0, \  S_1,S_2,J_3 \not=0$,\ \ \ \ \   i.e. $X_1,X_2=0, \ X_3 = 
e^{i \vp_3}
 ,\  \a=\vp_3  $

\noindent
In the first case we consider  string configurations with 
$\ Y_1,Y_2=0$ and $\x=t$ (so that $B_a=0$) 
 while in the second case we assume $X_1,X_2=0$ and thus $\a=\vp_3$ (and 
$C_a=0$).
In the first case we may  apply a boost, i.e. change the coordinates 
so that 
\be\la{uv} 
v\equiv \x=t \,,\qquad  u\equiv  \alpha -\x \,.
	\ee
 The \sm Lagrangian corresponding to the metric \rf{met} 
 then becomes  
\be
{L}=- \del^a v (\del_a  u   + C_a )
-\ha  ( \del_a u  + C_a)^2  -\ha   D^a U^*_i D_a U_i  \,,
\label{iio}
\ee
where $C_a=- i  U^*_i \del_a U_i$. We may choose the conformal gauge  and  supplement it with the condition 
\be 
v=\k \tau \,,
\ee
since this satisfies the equation of motion $\del^2 v=0$. 

As an aside, we note that in the more general case of non-zero $U_i$ and $V_i$ 
the equations of motion for $u$ and $v$ would be
\be 
\la{twe}
\del^2 u  + \del^a( C_a  - B_a) =0 \,,\qquad
 \del^2 v  + \del^a B_a=0 \,.  
\ee
In that case one could choose  $v=\k \tau$ only  when $\del^a B_a=0$.

Returning to the $V_0=1$, $V_1=V_2=0$ case, we may solve the conformal gauge constraints 
\ba
0&=& \del_0   v ( \del_1  u   + C_1)
 + \del_1   v ( \del_0  u   + C_0)
+ ( \del_0 u  + C_0) ( \del_1 u  + C_1)  \nonumber \\
&&
 + \ \ha (D_0 U^*_i D_1 U_i  + \mbox{c.c.})\,, \la{tyu}\\
0&=& 2\del_0   v ( \del_0  u   + C_0 )
 + 2\del_1   v ( \del_1  u   + C_1 )
+ ( \del_0 u  + C_0)^2 +  ( \del_1 u  + C_1)^2  \nonumber \\
&&
 + D_0 U^*_i D_0 U_i  
 +   D_1 U^*_i D_1 U_i\,, \la{eqr}
\ea
for $D_a u\equiv \del_a  u   + C_a$ and eliminate $u$ from the dynamics, 
getting an effective action for $U_i$ only. 
As follows from the constraints, to leading order (see also \ci{krt})
\be 
\k \approx \J= { 1 \ov \sqrt{ \tl}}  \ , \ \ \ \ \ \ \
J=\sql \J \ , \ \ \ \ \ \ \ \tl \equiv {\l \ov J^2} \ . \ee
Note also that 
$\del^a v (\del_a  u   + C_a )= - \k C_\tau+$total derivative, 
where $C_\tau$ is linear in time derivatives of $U_i$.
 To develop a ${ \k} \to \infty $ 
or $\tl \to 0$ expansion it is natural to re-scale \ci{krt}
 the $\tau$ coordinate by $\k$, introducing new time coordinate 
$\rt$ 
\be \la{resc}
\tau = \k \rt  \ , \ \ \ \  \ \ \ \ \ \ \ \ 
 t=\k \tau=  \k^2 \rt \approx \tl^{-1} \rt  \ . \ee 
Then  to the leading order in $\tl$ the action is 
\foot{Note that in the $\k \to \infty$ limit 
 the 
$\tau\tau$ component of the induced metric approaches zero, i.e.
the world sheet metric (in the  conformal gauge) 
 degenerates   (as was first 
 noted in a special
2-spin case in \ci{matt} and was further clarified and 
applied 
  in \ci{mik,mikk}).}
\be 
I = {\sql} \int d \tau  \inti \ L 
= J  \int d \rt   \inti \ \td L \ , \ \ \ 
\ \ \ \ \ \ \ \  \td L= \td L^{(0)} +   O(\tl)\ ,  \ee
\be \la{gh}
\td L^{(0)} =  C_0  -  \ha |D_1 U_i|^2  \ , \ \ \ \ \ \ \ \ \ \ \ 
C_0 \equiv  -
 i  U^*_i \del_\rt U_i  \ .   
\ee
In the two-spin case ($U_3,J_3=0$) considered in~\ci{kru,krt} 
the equations that follow from~\rf{gh} are the standard 
Landau-Lifshitz (LL) equations for a classical ferromagnet.~\foot{Let us note 
that 
in  the case when only one out of the three $U_i$'s is non-zero, 
 i.e. when we have   only one component of the angular momentum
 as in the point-like geodesic  case, then  $U_1=1, \ U_2,U_3=0, C_a=0$
 and so the action in \rf{iio} becomes trivial:
 $L =  -\del^a  t  \del_a  u  -\ha  ( \del_a u )^2 $.
 One  may study, however, quadratic 
  fluctuations near such a trajectory. The action that summarizes 
  them can be found by setting
 $ Y_0 = a e^{i t} ,  \ a^2 = 1 + |Y_s|^2  , \ 
 Y_s = \epsilon \td Y_s,$ and 
 $X_1 = r  e^{i (t + \epsilon^2 u) }  ,  \ r^2= 1 - |X_s|^2 , \ 
 X_s = \epsilon  \td X_s  \  (s=1,2)  .
 $
 Expanding in $\epsilon\to 0$ we get 
 from $ ds^2 =- a^2 dt^2 + r^2 d\a^2  + dr^2 + da^2 + 
 |dX_s|^2 + |dY_s|^2$:
 $
 L = -\ha \epsilon^2 \big[
  2 \del^a   t  \del_a  u -  (|\td X_s|^2 + |\td Y_s|^2) \del^a t
  \del_a t  
  + |\del_a X_s|^2 + |\del_a Y_s|^2\big]  + O(\epsilon^4) \ . 
 $
 Choosing the ``light-cone'' gauge $t= \k \tau$
 (and adding the fermionic part)  
 we are led \ci{ft1} 
  to the standard quadratic 
fluctuation  action of \ci{met}.}
To see this explicitly, define a unit vector~\cite{krt}
\be
n_i  ={\bf U}^\dagger\sigma_i {\bf U}\,,\ \ \ \ \ \ \ \ \
{\bf U}=\left(\begin{array}{c}U_1\\U_2\end{array}\right)\, , \ \ \ \   \ \ \ \ \ 
\ 
n_i n_i =1 \ , 
\ee
where $\s_i$ are Pauli matrices. Then~\rf{gh} becomes
\be\la{nnn}
\td L^{(0)} 
={\cal L}_{\mbox{\scriptsize WZ}}(n) -\frac{1}{8}(\p_1n_i)^2\,,
\ee
where
\be\la{wzz}
{\cal L}_{\mbox{\scriptsize WZ}}= C_0 = - 
\frac{1}{2}\int^1_0 d z \ \varepsilon^{ijk } n_i \p_z n_j  \p_0
n_k\,.
\ee
The corresponding  equations of motion are the usual 
Landau-Lifshitz equations
\be \la{LL} 
\p_0 n_i=\frac{1}{2}\e_{ijk} n_j\p^2_1 n_k\,.
\ee

In the second  case (which is also 
one-loop closed on the gauge theory side \ci{bs,bfst}) 
where $\a$ is a decoupled coordinate (like $t$ was  in the first case)
 it is  natural to set  instead  of \rf{uv}
\be
u\equiv \a \,,\qquad  v \equiv  \x -\a \,, 
\ee
 so that  
\be
{L}=  \del^a u (\del_a  v   + B_a )
+\ha  ( \del_a v  + B_a)^2  - \ha   D^*_a V^*_i D^a V^i 
 \,.
\label{iioo}
\ee
Then  choosing conformal gauge supplemented by 
\be 
u=\a = \J \tau \ , \ee
and dropping a total derivative term  we get 
\be
{L}=  - \J  B_\tau 
+\ha  ( \del_a v  + B_a)^2  - \ha   D^*_a V^*_i D^a V^i 
 \,. 
\label{ii}
\ee
The conformal gauge constraints  again determine $\del_a v  + B_a$ in terms
 of the ``transverse''  coordinates $V_i$. 
Rescaling the time coordinate as in \rf{resc}
\be
\la{esc}
 \tau = \J \rt  \ , \ \ \ \ \ 
  \ \  \a = \J  \tau=  \J^2 \rt =  \tl^{-1} \rt  \ , \ee 
we end up with a  systematic expansion of the sigma 
model action in powers of $\tl$  with 
the leading-order term in the effective Lagrangian being 
 $$ I = J  \int d \rt   \inti \ \td L \  , \ \ \ \ \ \ \ \ 
  \td L = \td  L^{(0)} +  O(\tl) \ , $$
\be \la{ghe}
\td L^{(0)}  =  - B_{0}  -  \ha D^*_1 V_i D_1 V^i   \  , \ \ \ \ \ \ \ \ \ \ \
 B_0\equiv  i V^*_i \p_\rt V^i \ . 
\ee
As $C_0$ in~\rf{gh}, here $B_{0}$, which is linear in time derivatives 
of $V_i$, plays the role of a WZ-type  term  in the action that  leads to 
generalized Landau-Lifshitz  equations.

It is instructive also  to present   the explicit  form of \rf{ghe} 
in terms of angular coordinates
in  the simplest non-trivial 
case of $S_1\not=0, \ S_2=0$. Then the  relevant part of the \adss 
metric is (cf. \rf{an},\rf{gl}) 
\be 
ds^2 = -  \cosh^2 \r \ dt^2 +  d\r^2 + \sinh^2 \r \ d \phi_1^2
\      +  \ 
 d \vp_3^2 \ . \ee 
Setting $ t= \x + \eta, \ \phi_1 =- \x + \eta, \ \ \a=\vp_3$  we get 
\be 
ds^2 =  - ( d \x + B)^2 +  d \a^2 
+   d\r^2 + \sinh^2 2 \r \ d \eta^2  \ , \ \ \ \ \ \ \ 
B= \cosh 2 \r \ d \eta \  . \ee 
The resulting leading term in the effective Lagrangian \rf{ghe} is then 
\be \la{kkk}
\td L^{(0)} = - \cosh 2 \r \ \del_0  \eta
 - \ha \big[ (\del_1\r)^2 + \sinh^2 2 \r \ (\del_1 \eta)^2\big]\ . \ee 
This is obviously an analytic continuation of the  Lagrangian 
in the $(J_1,J_2)$ (i.e. $SU(2)$) sector in~\ci{kru} ($\r \to i \psi, \ 
\eta \to - \beta$, see also~\ci{bfst}).
Introducing  an $SO(1,2)$ vector
\be \la{lla}
\el_0=\cosh 2 \rho\,,\qquad \el_1=\sinh 2\rho\ \cos 2\eta\,,\qquad 
\el_2=\sinh 2\rho\ \sin 2\eta\,,
\ee
which defines a hyperboloid 
\be\la{lol}
\eta^{ij}\el_i \el_j=-1\,,\ \ \ \ \ \ 
\ \ \qquad\eta^{ij}=\mbox{diag}(-1,1,1)\,,
\ee
we can re-write the Lagrangian \rf{kkk} in the same way as \rf{nnn}
\be
 \la{lll}
\td L^{(0)} = {\cal L}_{\mbox{\scriptsize WZ}}(l)   
-\frac{1}{8}\eta^{ij}\p_1\el_i\p_1\el_j \,, 
\ \ \ \ \ \ \ \ \ 
{\cal L}_{\mbox{\scriptsize WZ}}=-\frac{1}{2}\int^1_0 d z \ \varepsilon^{ijk} 
\el_i 
\p_z \el_j\p_0\el_k \ . 
\ee
This action is thus a direct ``analytic continuation'' of the $SU(2)$ action
\rf{nnn}.
The corresponding equation of motion is the 
analog of the Landau-Lifshitz equation \rf{LL} with $n_i \to \el_i$, 
i.e. with   $(-,+,+)$ signature.

\bigskip

\subsection{Matrix  Landau-Lifshitz equations}

Let us return to the first  3-spin  case of $S_1,S_2=0, \ J_i\not=0$  
and consider  another more explicit form of \rf{gh}
and the  generalized  LL equations that follow from it. 
Instead of the unit vector  $n_i$ we  may also 
use   an $SU(2)$ Lie algebra valued matrix $M$\  ($\a,\b=1,2$) 
\be
M_{\a\b}\equiv 2U^*_\a U_\b-\delta_{\a\b}\,,\qquad \ \  
n_i=\frac{1}{2}\sigma^{\a\b}_i M_{\a\b} \,.\label{defm}
\ee
The matrix $M_{\a\b}$ satisfies the relations 
\be\la{kon} 
{\rm Tr}\ {M}=0\,,\qquad M^\dagger=M\,,\qquad M^2=1\,,
\ee
with $M^\dagger$ being the hermitian conjugate of $M$.\footnote{These 
constraints reduce the 
number of independent real degrees of freedom that $M$ carries from eight to two 
- the same 
number as $U_1$ and $U_2$ have.}
In terms of $M$ the $CP^1$  Lagrangian ${\td L}^{(0)} (U_1,U_2,U_3=0)$
in \rf{gh} becomes
\be
{\td L}^{(0)} (M)={\cal L}_{\mbox{\scriptsize WZ}}(M)-
\frac{1}{16}\Tr(\p_1M\p_1M)\,,
\label{lssu2}
\ee
where 
\be
{\cal L}_{\mbox{\scriptsize WZ}}(M)=\frac{i}{8}\int^1_0  dz\  
 \Tr(M\left[\p_{z} M\,,\,\p_{0} M\right])\,.
\ee
The equation of motion for $M$ which follows 
from ${\td L}^{(0)}(M)$ is the matrix Landau-Lifshitz
 equation (see \ci{fadt})
 ~\footnote{We are  grateful to Gleb Arutyunov for 
bringing to our attention this form of the Landau-Lifshitz equation.}
\be
\p_0 M=-\frac{i}{4}\left[M,\p^2_1M\right]\,.\label{su2ll}
\ee
In the general  3-spin case ($U_3\neq 0$) we define an $SU(3)$ Lie algebra 
valued matrix $N$
($i,j=1,2,3$)
\be
N_{ij}\equiv 3U^*_i U_j-\delta_{ij}\,.
\ee
This matrix satisfies the following constraints (cf. \rf{kon}) 
\be\la{kkon} 
\Tr\ N =0\,,\qquad N^\dagger=N\,,\qquad N^2=N+2\,,
\ee
with the last constraint equivalent to $N^{-1}=\ha (N-1)$.\foot{One can shift 
$N$ 
by a constant to make it satisfy $N^2=1$ at the expense of 
the zero-trace condition.}
 The $CP^2$ Lagrangian in equation~\rf{gh} takes the form
\be
{\td L}^{(0)}(N)={\cal L}_{\mbox{\scriptsize WZ}}(N)
-\frac{1}{36}\Tr(\p_1N\p_1N)\, ,  
\label{lssu3}
\ee
where  the WZ term is 
\be
{\cal L}_{\mbox{\scriptsize WZ}}(N)=\frac{i}{18}\int^1_0  dz\   
\Tr(N\left[\p_{z} N\,,\,\p_{0} N\right])\,.
\ee
The  equation of motion for $N$ is the $SU(3)$ matrix Landau-Lifshitz equation
\be
\p_0N=-\frac{i}{6}\left[N\,,\,\p_1^2N\right]\,.\label{su3ll}
\ee
Similar expressions are found in the $(S,J)$ case \rf{ghe},\rf{kkk}.
In this sector 
we define an $SU(1,1)$ matrix ($\a,\b=0,1$) 
\be
L_{\a\b}\equiv 2V^*_\a V_\b+\eta_{\a\b}\,,\ \ \ \ \ \ \ \  \eta_{\a\b}
=\mbox{diag}(-1,1) 
\ .   \label{defl}
\ee
 Then the equations of motion 
corresponding to \rf{ghe}  can be written as 
\be
\p_0L=-\frac{i}{4}\left[L,\p^2_1L\right]\,,\label{su11ll}
\ee
where  matrix products are defined using $\eta^{\a\b}$.
Finally, for completness
let us  note that it is just as easy to 
write down the generalised Landau-Lifshitz equation for the
$SU(1,2)$ sector which corresponds 
to three  non-zero spins  $(S_1,S_2,J)$.
 This equationn is just an  analytic continuation 
of the equation~\eqref{su3ll}.

\section{From spin chains to  sigma models: $SU(3)$ sector}
The  Lagrangians \rf{gh} and \rf{ghe} 
 following from \rf{meta} are   two ``non-relativistic''
 sigma models with WZ-type terms  with the target spaces 
 $ CP^2= U(3)/[U(2)\times U(1)]$  and 
 ${\tilde {CP}}^2 = U(1,2)/[U(2)\times U(1)]$  respectively. 
The same expressions \rf{gh} and \rf{ghe} 
 will   appear  also on the SYM side as 
 the effective low-energy Lagrangians  of spin chains 
 associated to the one-loop dilatation operator in the respective sectors.
This 
generalizes the connection \ci{kru} 
between the 1-loop $SU(2)$ spin chain \sm and the 
leading-order limit of \adss \sm  from the  2-spin sector 
(operators Tr$(\Phi_1^{J_1} \Phi_2^{J_2})+\dots$) to more general
 chiral 3-spin sectors. 
  In the 2-spin $SU(2)$  sector the  symmetry group 
 is $G=U(2)$  and  one  factorizes  \ci{ran} over the stability  group of 
 a vacuum state, $H= U(1)\times U(1)$  getting  a 
 \sm on $G/H=  SU(2)/U(1) = CP^1$. 
 In  the  more general  case of the 3-spin $S^5$ sector 
 (operators Tr$(\Phi_1^{J_1} \Phi_2^{J_2} \Phi_3^{J_3})+\dots$) we have 
 $G= U(3)$  and  $H=U(2)\times U(1)$  so that  
 $G/H= CP^2$. 
\foot{One  may wonder if there is a similar limit that would allow 
 one to obtain a more general \sm that follows from the full 
 $SO(6)$ spin chain (i.e. $S^5$ \sm with a WZ term, 
 cf. \ci{ran}). It is not clear  if that is possible.
 The $SO(6)$ spin chain contain sectors of non-chiral operators 
 (e.g. Tr$(\Phi^* \Phi)^n ...$), and, a priori, 
   there is no reason 
 to expect to be able to match their 1-loop dimensions 
 with  classical string energies, even in the sector of ``long'' 
 scalar operators. However, there may be still another sub-sector 
 of states for which the matching may  be possible 
 -- pulsating string states \ci{emz}. Their
 role and  place in the present context remains to be understood
 but it seems that in this case the above 
 ``Hopf fibration'' parametrization 
 based on separation of one common direction in $S^5$ and a boost 
 is not a  useful one.}

In this section we find the continuum limit of the coherent state expectation value 
of the one loop $\cal N$=4 SYM dilatation 
operator in the $SU(3)$ sub-sector of operators Tr$(\Phi_1^{J_1} \Phi_2^{J_2} 
 \Phi_3^{J_3} )
+ ... $. 
Recall that the  one-loop dilatation operator in the scalar $SO(6)$ sector
is~\cite{mz1} 
\be
D_{_{SO(6)}}=\frac{\lambda}{(4 \pi)^2}\sum_{l=1}^{J}\left(K_{l,l+1}+2-
2P_{l,l+1}\right)\,,
\ee
where the trace, $K$, and permutation, $P$, operators act on $\R^6\otimes\R^6$ as
\be
K_{I_lI_{l+1}}^{J_lJ_{l+1}}=\delta_{I_lI_{l+1}}\delta^{J_lJ_{l+1}}\,,\qquad
P_{I_lI_{l+1}}^{J_lJ_{l+1}}=\delta_{I_l}^{J_{l+1}}\delta_{I_{l+1}}^{J_l}\,,
\ee
with $\R^6$ being the space of SYM  scalars  $\phi_I$   ($I=1,\dots,6$). 
The restrictions of $D_{_{SO(6)}}$ to $SU(2)$ and $SU(3)$ sectors 
 can be  easily  deduced  since these
 sub-sectors are traceless ($\Phi_{k} = \phi_k + i \phi_{k+3}$, \ $k=1,2,3$).
 In both sectors we have 
\be
D_{_{SU(3)}}=\frac{\lambda}{(4 \pi)^2}
\sum_{l=1}^{J}\left(2-2P_{l,l+1}\right)\,.\label{gsu}
\ee
In the $SU(2)$ subsector of $SU(3)$ sector 
 the permutation operator can be 
expressed in terms of $SU(2)$ 
generators ($S_i = \ha \s_i$ where $\s_i$ are 
Pauli matrices) 
\be
P_{l,l+1}=\frac{1}{2}+\frac{1}{2}\sum_{i=1}^3\sigma^i_l\sigma^i_{l+1}\,,
\ee
and so 
\be
D_{_{SU(2)}}=\frac{\lambda}{(4 \pi)^2}\sum_{l=1}^{J}\left(1-
\sum_{i=1}^3\sigma^i_l\sigma^i_{l+1}\right)\,.\label{gsu2}
\ee
Similarly, in the 
$SU(3)$ sector $P$   can be expressed in terms of the 
$SU(3)$ algebra generators $\lambda^r$, $r=1,\dots,8$  ($3 \times 3$ 
 Gell-Mann  matrices) 
\be
P_{l,l+1}=\frac{1}{3}+\frac{1}{2}\sum_{r=1}^8\lambda^r_l\lambda^r_{l+1}\,   . 
\ee
As a result, 
\be
D_{_{SU(3)}}=\frac{\lambda}{(4 \pi)^2}\sum_{l=1}^{J}\left(\frac{4}{3}-
\sum_{r=1}^8\lambda^r_l\lambda^r_{l+1}\right)\,.\label{gsu3}
\ee

\commentout{

\noindent More generally, it was shown in~\cite{bei} that the spin chain 
Hamiltonian related to the dilatation operator via
\be
D(g)=D_0+\frac{\lambda}{8\pi^2}H+ {\cal O} (\lambda^{2})\,,
\ee
can be written as a sum of nearest neighbour interactions of the form
\be
H_{12}=2h(J_{12})=\sum_j 2h(j)P_j\,,\label{hamproj}
\ee
where $J_{12}$ is the operator that measures total spin of the two fields, $P_j$ 
projects onto the $J_{12}=j$ sector, and $h(j)$ is the $j$-th harmonic number
\be
h(j)=\sum_{k=1}^j\frac{1}{k}=\Psi(j+1)-\Psi(1)\,,
\ee
with $\Psi(x)=\Gamma^\prime(x)/\Gamma(x)$.
}

In the following subsections we consider  the coherent state 
expectation values of these operators and determine the associated low-energy effective action.

\subsection{The $SU(2)$ subsector }

It is useful first to recall the derivation of the continuum
 limit of coherent state 
expectation
value of $D_{_{SU(2)}}$ (see \ci{fra,ran,kru} and 
references therein). With $S_i$ denoting the $SU(2)$ generators, 
   the coherent spin state can be defined by applying a rotation 
   $R(n)$ to the  highest-weight state oriented along 
   the third axis, which orients it along the unit vector $n_i$. 
   Equivalently~\ci{pere}, we may define it as
\be
\ket[n]\equiv e^{i(n'\times n_0)\cdot S}\ket[s,s]=e^{i(n'_yS_x-n'_xS_y)} 
\ket[s,s] \,,
\ee
where $n_0=(0,0,1)$, $(n')^2=1$ and $\ket[s,s]$ is a highest-weight state
\be
S_z\ket[s,s]=s\ket[s,s]\,,\qquad S^2\ket[s,s]=s(s+1)\ket[s,s]\,.
\ee
The key  property of the coherent state $\ket[n]$ is that it  satisfies
\be
\bra[n]S_i\ket[n]= s n_i\,, \ \ \ \ \ \ \ \ \   n_i n_i  =1 \ , 
\ee
where $n_i$ parametrizes the coset $SU(2)/U(1)$
$$
n_1=\frac{n_y'\sin\Delta}{\Delta}\,,\ \ \ \ \ \
n_2=\frac{-n_x'\sin\Delta}{\Delta}\,,\ \ \ \ 
n_3=\cos\Delta\,,\qquad\Delta=\sqrt{n_x'{}^2+n'_y{}^2}\, .
$$
In order to generalise to other groups it will be useful 
to work with the matrix $M$ in (\ref{defm}) 
 rather than the vector $n$.
The coherent state is then (here we consider the relevant case of 
 $s=\ha $)
\be
\ket[M]\equiv e^{i(a\sigma_1+b\sigma_2)}\ket[0]\,,
\ \ \ \ \ \ \ \  \ket[0] \equiv \ket[\ha,\ha] \ ,  \label{su2coh}
\ee
where  $a,b$ are angular variables. Then 
\be
\bra[M]\sigma_i\ket[M]=\bra[0]\sum_{j=1}^3a_{ij}\sigma_j\ket[0]=
\bra[0]a_{i3}\sigma_3\ket[0]=a_{i3}\,,
\ee
where
\be
a_{13}=-b\frac{\sin(2\Delta)}{\Delta}\,,\ \ \ 
a_{23}=a\frac{\sin(2\Delta)}{\Delta}\,,\ \ \
a_{33}=\cos(2\Delta)\,,\ \ \  \Delta=\sqrt{a^2+b^2}\ . 
\ee
The matrix $M$ is  
\be
M=\sum_{i=1}^3a_{i3}\sigma_i\,,\qquad\mbox{i.e. }\qquad
\bra[M]\sigma_i\ket[M] = a_{i3}=\frac{1}{2}\Tr(M\sigma_i)\, .
\ee
Explicitly, 
\be
M_{\a\b}=2U^*_\a U_\b -\delta_{\a\b}\,,\label{mu}
\ee
where
\be
U_1=\cos\Delta\  e^{-i\theta/2}\ ,\ \ \  U_2=\sin\Delta\  e^{i\theta/2}\ , \ \ \  
b+ ia\equiv  - \Delta e^{i\theta}\ , \ \ \ 
 \sum_{\a=1,2}|U_\a|^2=1\, 
\ee
i.e.  $U_\a$, and hence $M$  are coordinates on $SU(2)/U(1)$. 

Next, we may define a coherent  state for the whole spin chain as 
\be
\ket[M]\equiv\prod_{l=1}^J \ket[M_l]\,,
\ee 
where $\ket[M_l]$ is the coherent state 
~(\ref{su2coh}) at site 
$l$. Then 
\ba
\bra[M]D_{_{SU(2)}}\ket[M]&=&\frac{\lambda}{(4\pi)^2}
\sum_{l=1}^J\left[1-\frac{1}{4}\Tr(M_l\sigma^i)
\Tr(M_{l+1}\sigma^i)\right]\nonumber \\
&=&\frac{\lambda}{128\pi^2} \sum_{l=1}^J
\left[\Tr(M_l\sigma^i)-\Tr(M_{l+1}\sigma^i)\right]^2\ ,
 \ea
i.e.
\be \la{hop}
\bra[M]D_{_{SU(2)}}\ket[M]\ 
&\rightarrow&  J \inti\  \frac{{\tilde\lambda}}{32} 
  \left[\Tr(\p_1 M\sigma^i) + O({1\ov J} \del^2_1 M) \right]^2 \nonumber 
  \\
 &\rightarrow&  J \inti\ 
\frac{{\tilde\lambda}}{16} \Tr(\p_1  M\p_1 M)\,.
\label{hamsu2}
\ee
We have taken  a  continuum limit defining as in~\ci{kru,krt} 
($0<\sigma\le 2\pi$)
\be \la{nep}
M(\sigma_l)= M(\frac{2\pi l}{J})\equiv M_l
\,,\ \ \ \ \   M_{l+1}-M_l=
\frac{2\pi}{J}\p_1 M +{\cal O}(\frac{1}{J^2})\,,  \ \ \ 
  \p_1 \equiv {\del \ov \del \s}. 
\ee
We have assumed that  $J\to \infty$ for 
${\tilde\lambda}={\lambda\ov J^2}$=fixed. This ensures that terms subleading 
in $1/J$ drop out.
 We have also used the completness identity 
$
\left(\Tr(A\sigma^i)\right)^2=2\Tr A^2  $
for any traceless $2\times 2$ matrix $A$, and that $\Tr M_l^2=2$,
since $M$  can be written in the form~(\ref{mu}). 
The full action in the coherent state path integral contains also the WZ 
term that ensures the correct quantization conditions at each site. 
After the rescaling of the time $t\rightarrow \rt= \tl^{-1} t $ as in \ci{krt} 
the full action  becomes the same as in \rf{lssu2}
\be 
I = J \int d\rt \inti \ {\cal L}_{_{SU(2)}} \ , \ \ \ \ \ \ \ \ 
{\cal L}_{_{SU(2)}}
={\cal L}_{\mbox{\scriptsize WZ}}(M)-\frac{1}{16}\Tr(\p_1M\p_1M)\, .
\ee 
This implies, in particular, that 
the leading  order 
 $\tl$ term in the energy of the 2-spin string solutions 
agrees~\ci{kru,krt} with the one-loop term on the  SYM side 
computed in the same  $J\to \infty, \  
{\tilde\lambda}$=fixed  limit 
(which is thus a semiclassical
 limit on the spin chain side).

\subsection{The $SU(3)$ sector}

\noindent
With the $SU(3)$ generators in the fundamental 
representation 
chosen  as Gell-Mann matrices 
$\l^r$ 
(so that the  $SU(2)$ subgroup is generated by $\lambda^1,\lambda^2,\lambda^3$ 
and the Cartan generators are $\lambda^3$ and $\lambda^8$)
  we define the coherent state as 
\be
\ket[N]\equiv e^{i(a\lambda^4+b\lambda^5+
c\lambda^6+d\lambda^7)}\ket[0]\,,\label{su3coh}
\ee
where 
 $a,b,c,d$ are angular variables.
The state $\ket[0]$ will be chosen to  satisfy (see ~\cite{ran} and also 
 Appendix A)
\ba
\lambda^3\ket[0]=h_1\ket[0]\,,&\qquad&
\lambda^8\ket[0]=h_2\ket[0]\,,\label{hab}\\
\lambda^1\ket[0]=0\,,&\qquad&
\lambda^2\ket[0]=0\,,\label{hnonab}\\
\bra[0]\lambda^r\ket[0]=0\,,&\qquad&
 r=4,\dots,7\,. 
\ee
Equation~(\ref{hnonab}) implies that the constants $h_1,h_2$ are 
\be
h_1=0\,,\qquad \ \ \   h_2=-\frac{2}{\sqrt{3}}\,.
\ee
Then the  coherent state $\ket[N]$ satisfies
\be
\bra[N]\lambda_r\ket[N]=\bra[0]\sum_{m=1}^8b_{rm}\lambda^m\ket[0]=
\bra[0](b_{r3}\lambda^3+b_{r8})\lambda^8\ket[0]=-\frac{2}{\sqrt{3}}b_{r8}
\equiv \frac{2}{3}a_r\,.
\ee
Explicitly,  one finds for  $a_i$ 
\ba
\frac{2}{3} a_1&=&2(ac+bd)\frac{\sin^2\Delta}{\Delta^2}\,,\qquad\qquad\qquad\!\! 
\frac{2}{3} a_2=2(bc-ad)\frac{\sin^2\Delta}{\Delta^2}\,,\nonumber \\
\frac{2}{3} a_3&=&(a^2+b^2-c^2-d^2)\frac{\sin^2\Delta}{\Delta^2}\,,\qquad
\frac{2}{3} a_8=-\frac{1}{2\sqrt{3}}(1+3\cos 2\Delta)\,,\label{a1a8}\\
\frac{2}{3} a_4&=&\frac{b\sin 2\Delta}{\Delta}\,,\ \
\frac{2}{3} a_5=\frac{-a\sin 2\Delta}{\Delta}\,,\ \ 
\frac{2}{3} a_6=\frac{d\sin 2\Delta}{\Delta}\,,\ \ 
\frac{2}{3} a_7=\frac{-c\sin 2\Delta}{\Delta}\,,\nonumber
\ea
where now $\Delta=\sqrt{a^2+b^2+c^2+d^2}$.
The matrix $N$, labelling our coherent state $\ket[N]$ is 
\be\la{alg}
N=\sum_{r=1}^8a_r\lambda^r\,,\qquad\mbox{i.e. }\qquad
\bra[N]\lambda_r\ket[N]= { 2 \ov 3} a_r=\frac{1}{3}\Tr(N\lambda_r)\,,
\ee
where only four out of the eight real components $a_r$ are independent. 
This construction guarantees that the matrix $N$ has the following  
decomposition
($i,j=1,2,3$)
\be\la{nm}
N_{ij}\equiv 3U^*_iU_j-\delta_{ij}\,  ,\ \ \  \ \ \ \ \ \ \ \ \ 
\sum_{i=1}^3|U_i|^2=1\,,
\ee
where $U_i$ are defined without an overall phase. 
That means that 
  $U_i$,  and hence 
$N$,  are coordinates on $SU(3)/(SU(2)\times U(1))$ or $CP^2$.
 To see this explicitly let us note that any matrix $N$ from  $SU(3)$ algebra
 \rf{alg} 
 admits a representation  (\ref{nm}) in terms of $U_i$ if and only if
\ba
a_1^2+a_2^2&=&(1+a_3+{ a_8 \ov \sqrt 3})(1-a_3+{ a_8 \ov \sqrt 3})\,,\\
a_4^2+a_5^2&=&(1+a_3+{ a_8 \ov \sqrt 3})(1-2{ a_8 \ov \sqrt 3})\,,\\
a_6^2+a^7_2&=&(1-a_3+{ a_8 \ov \sqrt 3})(1-2{ a_8 \ov \sqrt 3})\,,  \ \ \ \ \ \ 
\sum_{r=1}^8 a_r^2=3\,. 
\ea
It is straightforward to check that the constants 
$a_r$ in equations~(\ref{a1a8}) satisfy these four equations.

We may then consider    the coherent   state 
for the whole spin chain
\be
\ket[N]\equiv\prod_{l=1}^J\ket[N_l]\,,
\ee 
where $\ket[N_l]$ are  defined in equation~(\ref{su3coh}).
 Computing the matrix element
\ba
\bra[N]D_{_{SU(3)}}\ket[N]&=&\frac{\lambda}{(4\pi)^2}\sum_{l=1}^J
\left[\frac{4}{3}-\frac{1}{9}\Tr(N_l\lambda^r)
\Tr(N_{l+1}\lambda^r)\right]\nonumber \\
&=&\frac{\lambda}{288\pi^2}\sum_{l=1}^J 
\bigg[\Tr(N_l\lambda^r)-\Tr(N_{l+1}\lambda^r)\bigg]^2\ , 
\ea
and taking the continuum limit with $J\to \infty, \ \tl$=fixed as in the $SU(2)$ 
sector we get,
\ba
\bra[N]D_{_{SU(3)}}\ket[N] \ 
&\rightarrow& 
J \inti\   \frac{{\tilde\lambda}}{72}\left[\Tr(\p_1 N\lambda^r)
 + O({1\ov J}\del^2_1 N)\right]^2 \nonumber\\ 
&\rightarrow& 
  J \inti\    \frac{{\tilde\lambda}}{36}\Tr(\p_1 N\p_1  N)\,.\la{hama} 
\ea
Here we have used that for any traceless $3\times 3$ matrix $A$
\be
\left(\Tr(A\lambda^r)\right)^2=2\Tr A^2\,,
\ee
 and that $\Tr N_l^2=6$,
 since $N$  can be written in the form~(\ref{nm}).  
Rescaling $t\rightarrow \rt= \tl^{-1} t$,   the total
 coherent state 
path integral action  becomes 
\be
I = J \int d\rt \inti \ {\cal L}_{_{SU(3)}} \ , \ \ \ \ \ \ \ \ \ 
{\cal L}_{_{SU(3)}}={\cal L}_{\mbox{\scriptsize WZ}}(N) 
-\frac{1}{36}\Tr(\p_1N\p_1N)\,.
\label{lgsu3}
\ee
The matrix $N$ satisfies the same constraints as in \rf{kkon}.
Again, the limit $J \to \infty$, \ $\tl=$fixed is  a  semiclassical
limit on the spin chain side, and the  classical action 
is thus identical  to
 the $CP^2$  sigma model Lagrangian 
 ${\td L}^{(0)}(N)$ in equation~(\ref{lssu3})
 or \rf{gh}.

As a result,  we have demonstrated the leading-order   equivalence 
(proposed in \ci{ft2,ft3,afrt} and checked  previously on particular examples 
in \ci{emz,kri})
  between
 the SYM  theory and the string theory in the  3-spin $SU(3)$ sector.
 This implies in particular the agreement between string energies and 
anomalous dimensions as well as a relation between integrable structures 
\ci{as,emz,engquist}.

\section{From spin chains to  sigma models:  $SL(2)$ sector}

Let us now 
 consider the  $SL(2,R)$ sector of the gauge theory  \ci{bei,bs}
  containing the operators 
 \be \la{po}
\Tr ({D}^S_{1+i2}\Phi^J) +\dots\,,
\ee
where $\Phi\equiv \Phi_3=\phi_{5}+i\phi_{6}$  and $D_{1+i2} = D_1 + i D_2$. 
This subsector is closed under renormalisation in perturbation theory, and is invariant 
under an  $SL(2)$ subalgebra of the superconformal  algebra \ci{bei}. 
The planar one-loop anomalous dilatation operator  is then 
found to be equivalent to the Hamiltonian of the XXX$_{-1/2}$ spin chain 
\ci{bei}. 
 The  spin chain has
 $J$ sites, with ${D}^{n_l}_{1+i2}\Phi$ at each site, 
 $ \sum^J_{l=1} n_l = S$, i.e. 
  the  ``spin variable''  at each site transforms in an infinite
 dimensional $s=-\frac{1}{2}$  representation of $SL(2)$. 
 
 This representation can be constructed by standard 
 oscillator methods. Introducing a pair  of creation and  annihilation 
 operators $a, a^\dg$ with $[a,a^\dg]=1$, one defines
 the $SL(2)$ generators as
 \be
\K_0=a^\dg a+\frac{1}{2}\,,\qquad \K_-=a\,,\qquad 
\K_+=a^\dg+a^\dg a^\dg a\,,\ \ \ \ \ \ \ \ \ \ 
\K_\pm\equiv \K_1\pm i \K_2\ . \label{sl2gens}
\ee
Then  $[a,a^\dg]=1$
implies the  $SL(2)$ commutation relations\foot{More generally, 
one may consider 
$
\K_0=a^\dg a - s, \ \  \K_-= a,  \ \ 
\K_+=- 2 s a^\dg+a^\dg a^\dg a, \  
$ 
with the $SU(2)$ case corresponding to $\K_+ \to - \K_+$.}
\be
[\K_+,\K_-]=-2\K_0\,,\qquad [\K_0,\K_\pm]=\pm \K_\pm\, . 
\ee
 The $SL(2)$ quadratic Casimir
\be
\K^2=\K^2_0 - \K^2_1 - \K^2_2 = 
 \K^2_0-\frac{1}{2}(\K_+\K_-+\K_-\K_+)
\ee
in this representaion is  equal to $-{1 \ov 4}$, 
i.e. to   $s(s+1)$ for  $s=-\ha$.
Defining the ``highest weight''  state $\ket[0]$    as
\be 
a \ket[0]=0 \ , \ \ \ \  \ \ \ 
\K_-\ket[0]=0\,,\qquad \K_0\ket[0]=\frac{1}{2}\,,
\ee
 we can then construct the representation by associating 
\be
\frac{1}{n!}(D_{1+i2})^n\Phi\ \to \  (a^\dg)^n \ket[0]
\,.
\ee
In  general~\cite{bei,bs}, the one-loop dilatation operator is 
\be
D  =\frac{2\lambda}{(4\pi)^2} H \,, \ \ \ \ \ \ \ \ 
H= \sum_{l=1}^J  H_{l,l+1} \ , 
\ee
where $H$ is a spin chain Hamiltonian  containing nearest neighbour interactions 
\be
H_{l,l+1}=2 h(\K_{l,l+1})=\sum_j  2h(j) P_j\,.\label{hamproj}
\ee
Above, $\K_{l,l+1}$ is the operator that measures total spin at the two sites
(i.e. $(\K_{l} + \K_{l+1})^2$ as the Casimir), 
$P_j$ projects
onto the $\K_{k,k+1}=j$ sector, and $h(j)$ is the $j$-th harmonic number
\be
h(j)\equiv  \sum_{k=1}^j\frac{1}{k}=\Psi(j+1)-\Psi(1)\,,
\ee
with $\Psi(x)=\Gamma^\prime(x)/\Gamma(x)$.
Explicitly, in the present case $H_{l,l+1}$ can be defined 
by its action on a generic two-site state~\ci{bei} as
\be \la{defn}
H_{l,l+1} (a^\dg_l)^k (a^\dg_{l+1})^{n-k}\ket[0,0] =
\sum_{p=0}^n
c_p (k,n)\  (a^\dg_l)^p (a^\dg_{l+1})^{n-p} \ket[0,0] \ , \ee
where
\be 
[c_p (k,n)]_{k=p} = h(k)+h(n-k)  \ , \ \ \ \ \ \ 
[c_p (k,n)]_{k\not=p}  = -\frac{1}{|k-p|} \,.\label{su11h}
\ee
As was found in \ci{bei}, this $H$    can be interpreted 
as a  Hamiltonian of the integrable XXX$_{-1/2}$ spin chain \ci{fad}.\foot{For 
 general
$s$,  one can define the Hamiltonian of the
{\it integrable }  XXX$_{s}$ spin chain 
as \ci{fad} 
$\sum_{l=1}^J [ \Psi ( \K_{l,l+1} + 1 ) -  \Psi (1) ] $
where 
$\K_{l,l+1}$ is defined in terms of  $SL(2)$ spin  variables  $\K^i_l$
(expressed in terms of the oscillators $a_l, a^\dg_l$ as explained 
in the previous footnote) 
  through 
$ \K_{l,l+1} (\K_{l,l+1} +1 ) = 2 s ( s+1) - 2 \eta_{ij} \K^i_l \K^j_{l+1}.$
Here  $\eta_{ij}=(-++)$ (in $SU(2)$ case 
$-\eta_{ij}=(+++)$).
}
The definition \rf{defn}   will be sufficient for our 
present aim of computing the expectation value of  $D_{_{SL(2)}}$
in the corresponding  $SL(2)$  coherent state. 

The  $SL(2)$ coherent state  is defined 
by applying a ``rotation'' to the highest weight sector 
that orients the third axis along a unit vector $\el_i$
in \rf{lol}. Equivalently, we may define it 
as \ci{pere}
\be
\ket[\el]=e^{i\tau (\sin\phi J_1-\cos\phi J_2)}\ket[0]\ , \ \ \ {\rm or} \ \ \ \
\ket[\el]=e^{\zeta J_+}e^{\eta J_0}e^{-{\bar\zeta} J_-}\ket[0]\,,\label{cohsl2}
\ee
where $\tau$ and $\phi$  are two real  ``angles'' related to one 
complex   parameter  $\zeta$ by 
\be
\zeta=\tanh\frac{|\tau|}{2}\,,\qquad  \eta\equiv \ln (1-|\zeta|^2)
 =-2\ln\cosh \frac{|\tau|}{2} \,.
\ee
The  second representation of the coherent state  is more useful  since 
$J_-\ket[0]=0\ ,\,  J_0\ket[0]=\frac{1}{2}$ and $J_+^k\ket[0]=\ k!\ a^\dg{}^k\ket[0]$
so that 
\be
\ket[\el]=(1-|\zeta^2|)^{1/2}\sum_{k=0}^\infty \zeta^k 
a^\dg{}^k\ket[0]\,.\label{su11coh}
\ee
The conjugate coherent state has to satisfy $\bra[\el]\ket[\el]=1$ and so is given by
\be
\bra[\el]=\bra[0]e^{{\bar\zeta} J_-}e^{-\eta J_0}e^{-\zeta J_+}
=\bra[0]e^{-\zeta J_+}e^{\eta J_0}e^{{\bar\zeta} J_-}\,.
\ee
Since $
\bra[0]J_+=0, \ \bra[0]J_-=\frac{1}{2}\bra[0], \ \bra[0]J_-^k=\bra[0]a^k$, 
 we may write it as
\be
\bra[\el]=(1-
|\zeta|^2)^{1/2}\sum_{k=0}^\infty\frac{{\bar\zeta}{}^k}{k!}\bra[0]a^k\,.
\ee
It is then straightforward to check the basic property  of the coherent state 
\be \la{lya}
\bra[\el] \K_i \ket[\el] = - \ha  \el_i \ , \ \ \ \ \ \  \ \ \ \ \ 
\eta^{ij} \el_i \el_j = -1 \ ,  \ee
where  the vector $\el_i$,  which parametrizes the hyperboloid $SU(1,1)/U(1)$,  is 
expressed in terms of $\zeta$ by  (cf. \rf{lla},\rf{lol}) 
\be \la{jh}
\el_0 = \frac{1+|\zeta|^2}{1-|\zeta|^2}\ , \ \ \ \ \ 
\el_1 = - \frac{ 2 \zeta   }{1-|\zeta|^2}\ , \ \ \ \ \ 
\el_2 = - \frac{ 2 \bar \zeta}{1-|\zeta|^2}\ . \ee 
Next, we may define the coherent state for the whole spin chain  
as the product of coherent states at each site, $
\ket[\el] \equiv\prod_{l=1}^J \ket[\el_l]$ 
and compute 
\be \la{sll}
\bra[\el]D_{_{SL(2)}}
\ket[\el]= {2 \l \ov ( 4 \pi)^2} 
\sum_{l=1}^J \bra[\el_{l,l+1}]H_{l,l+1}\ket[\el_{l,l+1}]\,.
\ee
The result of this computation (with details given in Appendix) 
is remarkably simple:
\be \la{simp}
\bra[\el]D_{_{SL(2)}}
\ket[\el]= {2 \l \ov ( 4 \pi)^2} 
\sum_{l=1}^J 
\ln  \frac{ 1 - \eta_{ij} \el^i_{l} \el^j_{l+1} }{2}   \ , \ee
or, equivalently, 
\be \la{imp}
\bra[\el]D_{_{SL(2)}}\ket[\el]= {2 \l \ov ( 4 \pi)^2} 
\sum_{l=1}^J 
\ \ln \big[ 1 + { 1 \ov 4}  \eta_{ij}( \el^i_{l} - \el^i_{l+1})
( \el^j_{l} - \el^j_{l+1}) \big] 
  \ . \ee
  It is interesting to note that 
  \rf{simp}   is the direct $(-++)$ signature 
  analog on the {\it classically}
   integrable  lattice Hamiltonian for the Heisenberg magnetic
   \ci{fadt}, which is explicitly given by
   $\sum_{l=1}^J \ln \big( \frac{ 1  + n^i_{l} n^i_{l+1} }{2}\big)$, 
   where $ n^i n^i =1$  (with  $n_3=\el_0$, $n_{1,2} = i \el_{1,2}$).
   \foot{In  contrast, in the $SU(2)$
    case with $s=1/2$   only the continuum (i.e. Landau-Lifshitz) 
    limit of the  
    coherent state expectation value 
    \rf{hop}, i.e. of 
     $\sum_{l=1}^J (1-  n^i_{l} n^i_{l+1})$,   is an integrable 
     classical system.
     In this case, the fact that both  $S^5$ spins 
     $(J_1,J_2)$ are large, implies that
     we need to consider large clusters of spins (that exist
     due to ferromagnetic attraction) which in turn effectively translates into a
     semiclassical limit.}

Note also that since we are interested in comparing to the 
semiclassical  string  case,  $S$  in \rf{po} 
as well as $J$  should be large, 
and that, combined with a ferromagnetic nature of the spin chain, 
 effectively corresponds to a low-energy  semiclassical limit of 
the chain.
 In general, starting 
with (see  footnote 10)
$H \sim \sum_{l=1}^J  \Psi ( \K_{l,l+1} + 1 )  $
where 
$ \K_{l,l+1} (\K_{l,l+1} +1 ) = 2 s ( s+1) - 2 \eta_{ij} \K^i_l \K^j_{l+1} $
and considering  a semiclassical limit in which $\K_{l,l+1}$ is large 
 \ci{beg}, one finds\foot{The
 $s=1/2$ case is special since here $ \K^i $ are proportional to 
 Pauli 
 matrices and thus any  function of $\K^i_l \K^j_{l+1}$ 
 reduces simply to  a linear function ($(\s_l \cdot \s_{l+1})^2
 = 3 -  2 \s_l \cdot \s_{l+1}$, etc.).} 
 $$H \sim \sum_{l=1}^J \ln \K_{l,l+1} \sim 
 \sum_{l=1}^J \ln  [ 2 s ( s+1) - 2  \eta_{ij} 
 \K^i_l \K^j_{l+1}]\ .$$ 
Then taking the coherent state expectation value assuming all 
correlators factorize and using that $ < \K^i_l > = s \el^i_l$
(see \rf{lya})   we indeed 
arrive at \rf{simp}.

The final step is to consider the  limit 
$J\to \infty$ with fixed $\tl = { \l\ov J^2}$
 which amounts to taking a 
  low-energy continuum limit of this ferromagnetic chain. 
As in the $SU(2)$ \rf{hamsu2},\rf{nep}    and $SU(3)$ \rf{hama}  sectors  here 
we set  
\be \la{rep}
\el(\sigma_l)= \el(\frac{2\pi l}{J})\equiv \el_l
\,,\ \ \ \ \   \el_{l+1}-\el_l=
\frac{2\pi}{J}\p_1 \el +{\cal O}(\frac{1}{J^2} \del^2_1 \el)\, ,  
\ee
where $\del_1\equiv \del_\s$ derivatives of $\el$ are assumed to 
be finite in the limit. 
Since we have only one power of $\l$ in \rf{simp}, 
   in expanding the logarithm we need to keep only the order $1\ov J^2$ term, 
   i.e. the term quadratic in first derivatives. 
 This leads to 
 \be \la{kou}
\bra[\el]D_{_{SL(2)}}\ket[\el] \to 
J\int_0^{2\pi}\frac{d\sigma}{2\pi}\ \bigg[\  \frac{{\tilde\lambda}}{8}
 \eta_{ij}\p_1 \el^i\p_1\el^j +
   {\cal O}(\frac{1}{J^2} \del^4_1) \bigg] \,.
\ea
In this way we  reproduce the  spatial derivative 
 term in the \sm action~(\ref{lll}). This  
 implies   the general agreement between the string 
 and SYM theories
 at leading order in $\tl$ in the $SL(2)$ sector and thus 
 generalizes the    previous results \ci{bfst}
  for  particular solutions.

\section{Comments on the non-holomorphic $SO(6)$ sector}

In the case  of more general sectors involving 
non-holomorphic
operators  it is presently not clear how to make a systematic comparison 
to (a limit of) the string sigma model. However, it may be of interest
to repeat the above discussion of the coherent state expectation values 
of the one-loop dilatation operator in the general $SO(6)$ scalar 
sector  of the operators 
$\Tr ( \phi_{i_1} ... \phi_{i_L})$, where $\phi_i$ are 6 real scalars  \ci{mz1}. 

In constructing coherent states for a group $G$ one identifies
a ``vacuum state'' $\ket[0]$ together 
with a (maximal) subgroup $H$ which leaves the vacuum 
state invariant (see Appendix A for
a more detailed discussion of the construction of coherent states). 
This implies that the non-Cartan elements of $H$ annihilate
$\ket[0]$. There are four maximal subgroups of $SO(6)$: $SO(5)$, 
$SO(3)\times SO(3)$, $SO(4)\times SO(2)$ and $SU(3)\times U(1)$.
As explained in Appendix A, 
in the case of the fundamental  representation of $SO(6)$ there are only two 
possible choices of $H$, which admit a suitable vacuum state:

(i) $H=SO(4)\times SO(2)$\ \  with 
$\ket[0]_{SO(4)\times SO(2)}=(0,0,0,0,1,i)$, 

(ii) $H=SO(5)$ \ \ with $\ket[0]_{SO(5)}=(0,0,0,0,0,1)$.

The first choice  corresponds to selecting the BPS operator
$\Tr  (\phi_5 + i \phi_6)^L$ as a  vacuum state.  In this case the coset space 
$G/H= SO(6)/[SO(4)\times SO(2)]$ is an 8-dimensional Hermitian symmetric space --
the 
Grassmann  manifold $G_{2,6}$ equivalent also to $SU(4)/S(U(2) \times U(2))
$.\foot{In general, the coset
$ SO(n)/[SO(q)\times SO(n-q)] = G_{q,n}(R) $
is a  real  Grassmann manifold  which consists of all $q$-dimensional linear
subspaces of $R^n$ \ci{wolf}.
In particular, $G_{1,2}= S^1,\  G_{2,4} = S^2 \times  S^2$.
$G_{2,n}$ spaces  are Hermitian symmetric spaces.}
In the second case the vacuum is represented by a non-chiral operator
 $\Tr (\phi_6)^L$, and $G/H= SO(6)/SO(5)$ is $S^5$. Below we discuss the two cases in turn. 
  
\subsection{The $SO(6)/[SO(4)\times SO(2)]$ case}

In this section we consider the coherent state $\ket[m]$ for $SO(6)/[SO(4)\times 
SO(2)]$. The
eight dimensional coset space $SO(6)/(SO(4)\times SO(2))$ is spanned
by $M_{i5}$ and $M_{i6}$ with $i=1,...,4$ ($M_{ij}$ are $SO(6)$ generators in fundamental representation)
and so the coherent state is
\be\la{you}
\ket[m]=\exp\left[\sum_{i=1}^4(a_iM_{i5}
+a_{i+4}M_{i6})\right]\ket[0]\,,
\ee
with $\ket[0]=(0,0,0,0,1,i)$. The state $\bra[m]$ is 
defined by a similar relation  so as to satisfy $\bra[m]\ket[m]=1$.
Introducing 
  an antisymmetric
   imaginary  $6 \times 6$ matrix $m^{ij}$ 
  ($(m^{ij})^* = - m^{ij} = m^{ji}$) 
\be
m^{ij}\equiv\bra[m]M^{ij}\ket[m]\,,\label{mij}
\ee
one can check that $\Tr\ m^2=2$ and $ m^3 =m$, 
 or   explicitly 
\be\la{coh}
\sum_{i,j=1}^6 m^{ij}m^{ji}=2\,, 
\ \ \ \ \ \ \ \ \ \ \ \sum_{k,l=1}^6
m^{ik}m^{kl}m^{lj}=m^{ij}\,.
\ee
Below we will be interested in $\bra[m]M^{ij}M^{kl}\ket[m]$. On symmetry 
grounds, 
\ba
\bra[m]M^{ij}M^{kl}\ket[m]&=&\frac{1}{2}(\delta^{il}m^{jk}-\delta^{ik}m^{jl}
-\delta^{jl}m^{ik}+\delta^{jk}m^{il})\nonumber\\
&-&\frac{1}{2}(\delta^{il}w^{jk}-\delta^{ik}w^{jl}-
\delta^{jl}w^{ik}+\delta^{jk}w^{il})\,,\label{mijmij}
\ea
where $w^{ij}$ is a symmetric matrix  with $\sum_i w^{ii}=2$ (quadratic 
Casimir condition). It is possible to show that $w^{ij}$ is
equal to  the square of  $m^{ij}$, 
\be
w^{ij}=\sum_{k=1}^6 m^{ik}m^{kj}\,.
\ee 
We  define the coherent state for the whole spin chain as the product 
of coherent states at each site 
$
\ket[m]\equiv\prod_{l=1}^L \ket[m_l],
$ 
where $L$ is the length of the chain.
The one-loop dilatation operator 
is proportional to the 
$SO(6)$ spin chain Hamiltonian which  is a sum of nearest-neighbour
 interactions
 $H_{l,l+1}$. In terms of the $SO(6)$ generators $(M^{ij}_{ab})_l$
at each site, it is given by~\cite{mz1}
\be
\label{so6ham}
D_{_{SO(6)}} = { \l \ov (4 \pi)^2}\sum_{l=1}^{L} H_{l,l+1} \ ,
 \ \ \ \ \ \ \ \ \ 
H_{l,l+1} =
M_l^{ij}M_{l+1}^{ij}-\frac{1}{16}(M_l^{ij}M_{l+1}^{ij})^2+\frac{9}{4}
\,.
\ee
Using the equations~(\ref{mij}) and~(\ref{mijmij}) we find that the 
expectation value of $H_{l,l+1}$ is
\ba
\bra[m]H_{l,l+1}\ket[m]&=&2+\frac{3}{4}m_l^{ij}m_{l+1}^{ij}
-\frac{1}{4}m_l^{ij}m_l^{jk}m_{l+1}^{kl}m_{l+1}^{li}\nonumber\\
&=&\frac{3}{8}\Tr [(m_l -m_{l+1})^2]+
\frac{1}{8}\Tr [(m^2_l - m^2_{l+1})^2] \,,
\ea
where $\Tr(m_1m_2)=m_1^{ij}m_2^{ji}$, etc., and we used 
 $\Tr\ m^2 =2 $ and $m^3=m$.

To take the continuum limit we  again assume
that $L \to \infty$ for fixed
$ \td \l \equiv { \l \ov L^2}$. 
Taylor-expanding as in \rf{nep}, 
$m(\s_{l+1})= m(\s_l) + { 2 \pi \ov L} \del_1 m+ ...$, 
we need to  keep only terms 
with at most two derivatives (higher derivative terms will be
suppressed by $1\ov L$)
\ba
\bra[m]D_{_{SO(6)}}
\ket[m]&\rightarrow&
\frac{\lambda}{(4\pi)^2} L \int_0^{2\pi}
\frac{d\sigma}{2\pi}\,\,\,
\   \left(\frac{2\pi}{L}\right)^2\Bigl[
\frac{3}{8}\Tr (\p_1 m)^2    \nonumber 
\\&&\qquad\qquad\qquad\qquad
+\ \frac{1}{4}\Tr(m^2(\p_1 m)^2)
+\frac{1}{4}\Tr(m\p_1 m)^2\Bigr] \\
&=& L\int_0^{2\pi}\frac{d\sigma}{2\pi}\ 
\frac{\tilde\lambda}{16} \Bigl[\frac{3}{2}\Tr(\p_1 m)^2
+\Tr(m^2(\p_1 m)^2)+\Tr(m\p_1 m)^2\Bigr]\,,\nonumber
\ea
i.e. (cf. \rf{hamsu2},\rf{hama},\rf{kou})
\be \la{gra}
\bra[m]D_{_{SO(6)}}
\ket[m]\rightarrow
L\int_0^{2\pi}\frac{d\sigma}{2\pi}\ 
\frac{\tilde\lambda}{8} \Bigl[\Tr(\p_1 m)^2 + 
{ 1 \ov 4} \Tr(m\p_1 m)^2\Bigr]\,.\la{kaak}
\ee
Here, we have used  the identity $\Tr(\p_1 m)^2 = 
2 \Tr(m^2(\p_1 m)^2)+\Tr(m\p_1 m)^2$, 
which follows from
$m^3 =m$.

It is possible to rewrite the  Grassmanian 
$G_{2,6}$  action corresponding to \rf{gra}
in an equivalent form  which is similar to 
the $CP^2$ action \rf{gh}  in the $SU(3)$ case (cf. \rf{lssu3}).
Introducing a complex unit vector $V^i$ ($i=1,...,6$) 
subject also to $V^i V^i=0$ 
one can show that a generic imaginary antisymmetric matrix $m^{ij}$ 
satisfying the constraints \rf{coh} may be written as
\foot{The constraint $m^3=m$ implies that $6 \times 6$ matrix 
 $m$ can have  eigenvalues equal to 
 $1,-1,1,-1,0,0$ or $1,-1,0,0,0,0$. The latter option is the only
 possibility in view of Tr $m^2=2$ condition. 
 Then $V^i$ and $V^i{}^*$  are eigenvectors for
 the eigenvalues 1 and -1.}
 \be \la{jok}
m^{ij} = V^i V^j{}^*- V^j V^i{}^*\ , \ \ \ \ \ 
V^i V^i{}^* =1 \ , \ \ \ \ V^i V^i=0\,. 
\ee
The constraints on $m^{ij}$ imply that it has 
$15-1 - 6=8$ independent parameters (which is the dimension of 
$G_{2,6}$). The constraints on $V^i$ leave $12-1-2=9$ 
real  parameters, but in addition $m^{ij}$ is invariant under 
$V^j \to e^{i\a} V^j$, so we may restrict $V^i$ to belong 
to $CP^5$, i.e.  $G_{2,6}$ is  a surface $V^2=0$ in 
$CP^5$. The components of $V^i$ can be explicitly expressed in
terms of 8 real parameters $a_n$ of the coherent state in 
\rf{you}. The  effective Lagrangian 
corresponding to \rf{gra} then takes the same 
 form as $CP^5$ analog of \rf{gh}, i.e. 
 (after rescaling time as in \rf{lssu3}; note that Tr$(m\del_a
 m)^2=0$)   
 \be   \la{ghp}
\td L^{(0)} =   -
 i  V^i{}^* \del_\rt V^i   -  \ha |D_1 V^i|^2
 = - i V^i{}^*  \del_\rt V^i  -  \ha ( |\del_1 V^i|^2  
 - |V^i{}^* \del_1 V^i|^2)   \ ,    
\ee 
with the  constraint $V^2=0$
(a similar action with an additional constraint 
$V^i{}^*  \del_1  V^i =0$ was found on the string 
side in \ci{mikk}). The $SU(3)$ sector is the special case when 
$V^{a}= i V^{a+3}\equiv { 1 \ov \sqrt 2} U^a$,\ $a=1,2,3$, 
so that $|U^a|^2=1$, i.e. $U^a$ belongs to $CP^2$
and  \rf{ghp} reduces to \rf{gh} or \rf{lssu3}.

The above  coherent state  description  thus 
does not capture states 
with large ``extensive'' 
one-loop shift of the dimension 
 $E= L + c_1 \l  L + ...$ \ci{mz1}, but should instead
  describe 
the most general near-BPS sector of 
 semiclassical string states (including pulsating strings
 \ci{emz}) on $S^5$ for which one gets again a regular scaling limit, 
 that is a regular dependence of the
  one-loop correction on  $\td \lambda$. 
 
 The precise relation to string theory (in particular, to 
 the sector of pulsating string states) 
 still remains to be understood (for an interesting approach 
 in this direction  see  \ci{mikk}).  

\subsection{The $SO(6)/SO(5)$ case}

In this section we consider the coherent state 
$\ket[v]$ for $SO(6)/SO(5)$. The
5-dimensional coset space $SO(6)/SO(5)$ is spanned by $M_{i6}$ with 
$i=1,\dots,5$, and so 
the coherent states are given by
\be
\ket[v]=\exp\left[\sum_{i=1}^5a_iM_{i6}\right]\ket[0]\,,\qquad
\bra[v]=\bra[0]\exp\left[-\sum_{i=1}^5a_iM_{i6}\right]\,.
\ee
As discussed in Appendix A, here  the vacuum state is $\ket[0]=(0,0,0,0,0,1)$.
Since $\bra[v]$ is the transpose of $\ket[v]$, the components of the two are 
indentical. This, together with the fact that the $M^{ij}$ are anti-symmetric 
matrices, implies that
\be
\bra[v]M^{ij}\ket[v]\equiv 0\,,\label{vij}
\ee
for all $i,j=1,\dots,6$. Next, on symmetry grounds, one has
\be
\bra[v]M^{ij}M^{kl}\ket[v]=\delta^{il}v^{jk}-\delta^{ik}v^{jl}-
\delta^{jl}v^{ik}+\delta^{jk}v^{il}\,,\label{vijvij}
\ee
where $v^{ij}$ is a symmetric tensor with $\sum_i v^{ii}=1$ (quadratic 
Casimir condition in fundamental representation).  One can show that 
\be
v^{ij}=v^iv^j\,,
\ee
with 
\be
v^i=\frac{a_i\sin\Delta}{\Delta}\,,\ \ \  i=1,\dots,5\,,\ \ \mbox{and }\ \ 
v^6=\cos\Delta\,,\ \ \ \Delta=\sqrt{\Sigma_{i=1}^5\,\,a_i^2}\ . 
\ea
 The quadratic Casimir condition 
now reduces to
\be
\sum_{i=1}^6(v^i)^2=1\,.
\ee
In other words, $v^i$ are coordinates on $S^5$.

\noindent Defining  the state of  the whole spin chain as the 
product 
of coherent states at each site 
$
\ket[v]\equiv\prod_{l=1}^L \ket[V_l],
$ 
and 
using equations~(\ref{vij}) and~(\ref{vijvij}) we find that the 
expectation value of $H_{l,l+1}$ in (\ref{so6ham}) is
\ba
\bra[v]H_{l,l+1}\ket[v]=2-(v_l^iv_{l+1}^i)^2
=1+(v^i_l-v^i_{l+1})^2-\frac{1}{4}(v^i_l-v^i_{l+1})^4\,.
\ea
If we take the limit $L \to \infty$ 
we may consider   the continuum limit   and drop
higher derivative terms  
(assuming that derivatives over $\s$ are fixed in the large $L$ limit).
 Then (cf. \rf{hama},\rf{kou},\rf{kaak})
\ba
\bra[v]D_{_{SO(6)}}
\ket[v]&\rightarrow&
\frac{\lambda}{(4\pi)^2} L \int_0^{2\pi}
\frac{d\sigma}{2\pi}\,\,\,
\  \left[ 1+\left(\frac{2\pi}{L}\right)^2(\del_1 v^i)^2\right]\,.
\label{contso6so5}
\ea
The presence of the first 
$v_i$-independent term   implies that here 
 we do not  get a regular expansion in $\tl = {\l\ov L^2}$. 
The
coherent state expectation value thus captures the 
 large ``extensive'' one-loop shift of dimensions
 $E= L + c_1 \l  L + ...$ of 
the type described in~\ci{mz1} which should be characteristic 
of some oscillator string states. In these cases one expects 
the full expression for the energy/dimension  to contain functions of 
coupling 
interpolating from weak to strong coupling:
$E_{L\to \infty} = L + f_1 (\l) L +  L^{-1} f_2(\l; n_s) + ...$, 
where $f_k(\l\to 0) = c_k \l + b_k \l^2 + ... $
and $n_s$ stand for other (oscillator and/or spin) 
quantum numbers encoded in $v_i$. 
One may expect that for states with large quantum numbers  
$f_k(\l \to \infty) = a_k \sqrt{\l} + d_k + ...$, 
as,  for example, for a single $S^5$ spin string states \ci{gkp}.

\section{Conclusions}

In this paper we have demonstrated, 
in the large spin limit and to 
 leading order in the coupling $\tl= {\l \ov J^2}$,
 the equivalence of the 
 string theory on $AdS_5\times S^5$ and 
$N=4$ $SU(N)$ SYM gauge theory in the chiral 
$SU(3)$ and $SL(2)$ sectors 
of states.  We have developed an  expansion of 
the string sigma model Lagrangian, whose leading order term
was shown to describe   generalizations of the 
Landau-Lifshitz equations for a classical ferromagnet. 
On the SYM side, we have computed 
a coherent state expectation value 
 of 
the  corresponding 
spin-chain Hamiltonian which encodes the one-loop dilatation operator 
of the theory. In the thermodynamic limit, 
the resulting coherent-state 
sigma model matched exactly with the leading order action obtained from the 
string sigma model Lagrangian.
In this way we have generalised the recent results of~\cite{kru,krt} on the 
$SU(2)$ sector.
The matching of the two sigma model Lagrangians
implies a general agreement 
(at leading order in $\tl$) 
between the 
string energies and SYM anomalous dimensions  as well
as  matching of integrable structures, thus generalising previous 
results in these sectors~\cite{ft1,afrt,bfst,emz,engquist,as,kri} 
for particular solutions.

\noindent 
While the matching of  
various chiral sectors at leading order in the coupling now
seems to be well understood, such an understanding of the more general $SO(6)$ 
non-chiral sector of operators is still missing.
As a  step in this direction we computed the continuum limit of the $SO(6)$ spin chain  
Hamiltonian of \cite{mz1}. In doing this   the choice
of a vacuum state becomes  important. 
One can choose the vacuum to correspond to the BPS operator $\Tr \Phi^L$, in 
which case the resulting sigma model has as its target space 
the 8-dimensional 
 Grassmann manifold  $SO(6)/[SO(4)\times SO(2)]$. Then there are no 
 non-derivative  leading 
order corrections
to the $E=L$ relation, as
 should be the case for  a sector with a BPS ground
state. 
We expect this sigma model to be related to a gauge theory sector with a regular 
${\lambda\ov L^2}$  expansion of anomalous dimensions 
(such as the pulsating string solutions 
of~\cite{emz}). 
Since the target  space is 8-dimensional, it is not 
immediately clear how to relate it to a subsector of the 
string sigma model (see, however, \ci{mikk}). 
On the other hand, one can choose the vacuum 
to be represented by a real non-BPS operator 
$\Tr\phi^L$ whose dimension  receives ``extensive'' (order $L$) 
 leading order 
correction.  As was shown in section 5.2, this behaviour,
typical of more  general $SO(6)$ states~\cite{mz1},
is indeed captured by the continuum limit of the corresponding coherent-
state
expectation value of the $SO(6)$ spin chain Hamiltonian.
 Furthermore,
in this case the target manifold is indeed $S^5$ suggesting 
that a
direct relation to
the \adss string sigma model may be possible.

\bigskip
\bigskip

 While this paper was  nearing completion  
there appeared an interesting preprint~\ci{lop}  which also 
extends the $SU(2)$ 
result of~\ci{kru} to the $SU(3)$ sector. Our approach is somewhat
different (in particular,  we use a more covariant parametrization,
and exhibit a  
relation to the matrix Landau-Lifshitz equation)
but the final expressions for the actions in that sector agree
once expressed in terms of the same  coordinates.

\section*{Acknowledgments }
We are grateful to  G. Arutyunov, N. Beisert, M. Kruczenski, 
N. Nekrasov, Yu. Obukhov,  A. Ryzhov, F. Smirnov,  M. Staudacher and K. Zarembo 
  for useful discussions, suggestions  and comments.
The work of B.S. was supported by a Marie Curie Fellowship. 
The  work of A.T. was supported  by DOE
grant DE-FG02-91ER40690, the INTAS contract 03-51-6346
and RS Wolfson award.  
B.S. would like also to thank Gleb Arutyunov and 
the Albert Einstein Institute in Potsdam for hospitality during the 
initial stages of this project.

\setcounter{footnote}{0}
\setcounter{section}{0}

\appendix{General definition of  coherent states and $SO(6)$ case}

In this appendix we briefly review the definition of coherent states 
following  closely~\ci{pere}.
Given a semisimple group $G$ in the Cartan basis 
 $(\rH_i,\rE_\a,\rE_{-\a})$ 
($[\rH_i,\rH_j]=0, \ [\rH_i,\rE_\a ]=\a_i \rE_\a  , \ 
[\rE_\a, \rE_{-\a} ]= \a^i \rH_i , \ 
[\rE_\a, \rE_{\b} ]= N_{\a\b} \rE_{\a+\b}$) 
whose interpretation will be 
 a symmetry group of a quantum  Hamiltonian (acting 
in a  unitary irreducible representation $\L$ on the Hilbert 
space $V_\L$)
one may define a set of coherent states
 by choosing a particular state 
$\ket[0]$ (with $ \bra[0]\ket[0]=1$)  in $V_\L$
and acting on it  by the elements of $G$. A subroup $H$ of $G$ that
leaves 
$\ket[0]$ invariant up to a phase 
($ \L(h)\ket[0] = e^{i\phi(h)} \ket[0]$) 
 is called maximum stability subgroup.
One may then define the coset space $G/H$ elements  of  which 
($g = \omega h, \ h \in H, \ \omega \in G/H$, \ $\L(g)
=\L(\omega)\L(h)$) 
will parametrize the coherent states, 
$\ket[\omega,\L] = \L(\omega) \ket[0]$.

This definition depends on a choice of group $G$, its
representation 
$\L$ and the vector   $\ket[0]$. 
It is natural to assume also 
that $\ket[0]$ is an eigenstate 
 of the Hamiltonian $H$, e.g., a ground state. 
 For a unitary representation $\L$ we may choose 
 $\rH_i^\dagger=\rH_i  ,\ \rE_\a^\dagger = \rE_{-\a}$
 and select   $\ket[0]$ to be the highest-weight vector
  of the 
 representation  $\L$, i.e. demand that it is annihilated by
 ``raising'' generators  and is an eigen-state of the Cartan 
 generators: 
 (i) $ \rE_\a \ket[0]=0$ for all positive roots $\a$; \ \ 
 (ii) $\rH_i  \ket[0] = h_i  \ket[0]$.
 In addition, we may demand that $ \ket[0]$ is annihilated also 
 by some ``lowering'' generators, i.e. 
 (iii)  $ \rE_{-\b} \ket[0]=0$ for {\it  some}  negative 
  roots $\b$; the remaining negative roots will be denoted by $\g$. 
  Then  the  coherent states are given by   
  \be 
\ket[\omega,\L] = {\rm exp}\left[ \sum_\g (w_\g  E_{-\g} -
  w^*_\g  E_{\g} )\right]\ 
  \ket[0]\ , 
\ee
  where $\g$ are the negative roots for which $\rE_\g \ket[0]
  \not=0$.  $w_\g$ may be interpreted as coordinates 
  on $G/H$ where $H$ is generated by $(\rH_i,\rE_\a,\rE_{-\b})$.
  
  For example, in the case of $G=SU(3)$ 
  with the  Cartan basis 
  $$ (\rH_1,\rH_2, \rE_\a,\rE_{\b}, \rE_{\a+\b},
  \rE_{-\a},\rE_{-\b}, \rE_{-\a-\b})  $$
  and 
  with $\ket[0]$ 
  being the highest-weight of the fundamental representation
  (discussed in section 3.2), 
  i.e. $\rE_{-\b}\ket[0]=0$, $\rE_{-\a}\ket[0]\not=0$, 
  $\rE_{-\a-\b}\ket[0]\not=0$,
  the subgroup $H$ is generated 
  by $(\rH_1,\rH_2, \rE_\b,\rE_{-\b})$, i.e. is $SU(2)\times U(1)$
  and $G/H = SU(3)/(SU(2)\times U(1))= CP^2$.

  In section 5  we are  interested  in the case of the 
  fundamental representation of $SO(6)$.
  Then  we may write $\ket[0]$ as a linear superposition of 
  the highest-weight states of the fundamental
  irreducible  representation invariant under a maximal subgroup $H$ 
  of $SO(6)$.  
There are four such maximal subgroups for $SO(6)$: $SO(5)$, 
$SO(3)\times SO(3)$, $SO(4)\times SO(2)$ and $SU(3)\times U(1)$.
The first (last) two contain two (three) Cartan 
generators of $SO(6)$. It is easy to write down the generators of 
the first three subgroups in terms of the 15 generators $M_{ij}= (  
\delta^a_i\delta^b_j-\delta^a_j\delta^b_i)$ 
in the  fundamental representation of $SO(6)$: 
\ba
SO(5)&=&\left\{M_{ij}:i,j=1,\dots,5\right\}\,,\nonumber \\
SO(3)\times
SO(3)&=&\left\{M_{ij}:i,j=1,2,3\right\}\cup
\left\{M_{ij}:i,j=4,5,6\right\}\,,\nonumber \\
SO(4)\times 
SO(2)&=&\left\{M_{ij}:i,j=1,\dots,4\right\}\cup\left\{M_{56}\right\}\,.
\ea
To obtain the $SU(3)\times U(1)$ subgroup it is convenient to  use the Cartan 
basis 
of  the $SO(6)$ algebra. 
 The 3 Cartan generators 
may be chosen as  (linear combinations of) $M_{12}$, $M_{34}$, $M_{56}$.
The non-Cartan 
elements are (up to normalisation constants)
\ba
\rE_{\alpha}&=&M_{13}-iM_{14}-iM_{23}-M_{24}\,,\qquad
\rE_{\beta}=M_{15}-iM_{16}+iM_{25}+M_{26}\,,\nonumber \\
\rE_{\gamma}&=&M_{35}-iM_{36}+iM_{45}+M_{46}\,,\qquad
\rE_{\alpha+\beta}=M_{35}-iM_{36}-iM_{45}-M_{46}\,,\nonumber \\
\rE_{\alpha+\gamma}&=&M_{15}-iM_{16}-iM_{25}-M_{26}\,,\qquad
\rE_{\gamma-\beta}=M_{13}+iM_{14}-iM_{23}+M_{24}\,,\nonumber
\ea
with the negative roots $\rE_{-\alpha}=(\rE_{-\alpha})^*$, {\it etc}. 
Then $SU(3)\times U(1)$ is
generated by the three Cartan generators together with
\be 
\rE_{\pm\beta}\,,\rE_{\pm\gamma}\,,\ \ \ \ \ \ \ 
\rE_{\pm(\gamma-\beta)}\,.
\label{su3noncart}
\ee
Explicitly,  the Cartan generators associated to $\beta$ and $\gamma$ 
are proportional to
\be
\rH_\beta=M_{12}-M_{34}\,,\qquad \rH_\gamma=M_{12}+M_{34}-2M_{56}\,,
\ee
with the $U(1)$ generator $\rH_{U(1)}$ proportional
 to $M_{12}+M_{34}+M_{56}$. 

\noindent Let us now identify suitable vacua. For $H=SO(5)$ we have
\be
\ket[0]_{SO(5)}=(0,0,0,0,0,1)\,,
\ee
since then $M_{ij}\ket[0]_{SO(5)}=0$ for $i,j=1,\dots,5$. 
For $H=SO(4)\times SO(2)$ we have instead 
\be
\ket[0]_{SO(4)\times SO(2)}=(0,0,0,0,1,i)\, . 
\ee
In the case of  $H=SO(3)\times SO(3)$ subgroup the Cartan generators
may be chosen as  $M_{12}$ and $M_{56}$ which have four eigenvectors
$$
(i,1,0,0,0,0)\,,\ \ \ 
(-i,1,0,0,0,0)\,,\ \ \ 
(0,0,0,0,i,1)\,,\ \ \ 
(0,0,0,0,-i,1)\,,\ \ \ 
$$
whose eigenvalues are $(\mp i,0)$ and $(0,\pm i)$,  and two eigenvectors 
$$
(0,0,1,0,0,0)\ ,\ \ \ \ \ \ \ \ \ \  \ (0,0,0,1,0,0)$$ 
with zero  eigenvalues.
 Clearly,  the first four eigenvectors are not annihilated by the
non-Cartan part of $SO(3)\times SO(3)$ (that is by 
$M_{13}$, $M_{23}$, $M_{45}$ and $M_{46}$). It
is also easy to check that no linear combination
of the latter two eigenvectors is annihilated by the non-Cartan part of 
$SO(3)\times SO(3)$.
We conclude  that,   in the fundamental representation of $SO(6)$,
 there is no $SO(3)\times SO(3)$ invariant vacuum state.
 
Similarily  for $H=SU(3)\times U(1)$ there is no invariant vacuum state,
in the anti-symmetric (six-dimensional) representation of $SU(4)$.
 The eigenvectors of the Cartan
generators are:
\be
(\mp i,1,0,0,0,0)\,,\qquad (0,0,\pm i,1,0,0)\,,\qquad
(0,0,0,0,\pm i,1)\,.
\ee
Their $(\rH_\beta,\rH_\gamma,\rH_{U(1)})$ weights are $\pm(i,i,i)$, $\pm (i,-i,-i)$ and
$\pm (0,2i,-i)$, respectively. It is not difficult to see that none of the above six states is annihilated by all of the non-Cartan 
generators~(\ref{su3noncart}). 
This conclusion   is  clearly 
representation dependent; for example, 
 in the fundamental (four-dimensional) representation of
$SU(4)$ we can easily find a suitable ground state: this is 
just $(0,0,0,1)$.

\setcounter{equation}{0}
\setcounter{footnote}{0}
\appendix{Coherent state expectation value of $SL(2)$
spin chain  Hamiltonian }

Our aim here is to compute 
 the expectation value 
 $ \bra[\el_{l,l+1}]H_{l,l+1}\ket[\el_{l,l+1}]$ in 
 \rf{sll}.  Let us set for notational simplicity $l=1,\ l+1=2$ 
 and define 
\be
{\cal M}\equiv
\frac{\bra[\el_{12}]H_{12}\ket[\el_{12}]}{(1-|\zeta_1|^2)(1-|\zeta_2|^2)}\, . 
\ee
One finds using \rf{defn} 
\ba
{\cal M}
&=&\sum_{n_1,n_2,m_1,m_2=0}^\infty
\frac{\zeta_1^{n_1}\zeta_2^{n_2}{\bar\zeta}_1^{m_1}{\bar\zeta}_2^{m_2}}{m_1!m_2!
}
\bra[0]a_1^{m_1}a_2^{m_2}H_{12}a_1^\dg{}^{n_1}a_2^\dg{}^{n_2}\ket[0]\nonumber \\
&=&\sum_{m_1,m_2=0}^\infty\sum_{n=0}^\infty\sum_{k=0}^n
\frac{\zeta_1^k\zeta_2^{n-k}{\bar\zeta}_1^{m_1}{\bar\zeta}_2^{m_2}}{m_1!m_2!}
\bra[0]a_1^{m_1}a_2^{m_2}H_{12}a_1^\dg{}^ka_2^\dg{}^{n-k}\ket[0]\nonumber \\
&=&\sum_{m_1,m_2=0}^\infty\sum_{n=0}^\infty\sum_{k,l=0}^n
\frac{\zeta_1^k\zeta_2^{n-k}{\bar\zeta}_1^{m_1}{\bar\zeta}_2^{m_2}}{m_1!m_2!}
c_l (k,n)
\bra[0]a_1^{m_1}a_2^{m_2}H_{12}a_1^\dg{}^la_2^\dg{}^{n-l}\ket[0]\nonumber \\
&=&\sum_{n=0}^\infty\sum_{k,l=0}^n
\zeta_1^k\zeta_2^{n-k}{\bar\zeta}_1^l{\bar\zeta}_2^{n-l} c_l (k,n)\ , 
\ea
i.e. 
\be 
{\cal M}= {\cal M}_1+{\cal M}_2+\bar {\cal M}_2 \ee
where 
\be {\cal M}_1 = \bar {\cal M}_1
=\sum_{n=0}^\infty\sum_{k=0}^n
|\zeta_1|^{2k}|\zeta_2|^{2n-2k}[h(k)+h(n-k)]\ , \ee
\be {\cal M}_2= 
-\sum_{n=0}^\infty\sum_{k=0}^n\sum_{l=0}^{k-1}\zeta_1^k\zeta_2^{n-k}
{\bar\zeta}_1^l{\bar\zeta}_2^{n-l}\frac{1}{k-l} \ . 
\ea
Computing  these terms
explicitly we get 
\ba
{\cal M}_1&=&\sum_{n=0}^\infty\sum_{k=0}^n
|\zeta_1|^{2k}|\zeta_2|^{2n-2k}[h(k)+h(n-k)]\nonumber \\
&=&\sum_{m_1,m_2=0}^\infty|\zeta_1|^{2m_1}|\zeta_2|^{2m_2}[h(m_1)+h(m_2)]
\nonumber \\
&=&\frac{1}{1-|\zeta_1|^2}\sum_{m_2=0}^\infty|\zeta_2|^{2m_2}h(m_2)
+\frac{1}{1-|\zeta_2|^2}\sum_{m_1=0}^\infty|\zeta_1|^{2m_1}h(m_1)\nonumber \\
&=&-\frac{\ln(1-|\zeta_2|^2)+\ln(1-|\zeta_1|^2)}{(1-|\zeta_1|^2)(1-|\zeta_2|^2)}
\ , 
\ea
\ba
{\cal M}_2&=&-\sum_{n=0}^\infty\sum_{k=0}^n\sum_{l=0}^{k-1}\zeta_1^k\zeta_2^{n-
k}
{\bar\zeta}_1^l{\bar\zeta}_2^{n-l}\frac{1}{k-l}\nonumber \\
&=&-\sum_{m_1,m_2=0}^\infty\sum_{l=0}^{m_1-1}\zeta_1^{m_1}\zeta_2^{m_2}
{\bar\zeta}_1^l{\bar\zeta}_2^{m_1+m_2-l}\frac{1}{m_1-l}\nonumber \\
&=&-\frac{1}{1-|\zeta_2|^2}\sum_{m_1=0}^\infty\sum_{l=0}^{m_1-1}\zeta_1^{m_1}
{\bar\zeta}_1^l{\bar\zeta}_2^{m_1-l}\frac{1}{m_1-l}\nonumber \\
&=&-\frac{1}{1-|\zeta_2|^2}\sum_{n_1,n_2=0}^\infty\zeta_1^{n_1+n_2+1}
{\bar\zeta}_1^{n_1}{\bar\zeta}_2^{n_2+1}\frac{1}{n_2+1}\nonumber \\
&=&-\frac{1}{(1-|\zeta_2|^2)(1-
|\zeta_1|^2)}\sum_{n_2=0}^\infty(\zeta_1{\bar\zeta_2})^{n_2+1}\frac{1}{n_2+1}
\nonumber \\
&=&\frac{\ln(1-\zeta_1{\bar\zeta}_2)}{(1-|\zeta_2|^2)(1-|\zeta_1|^2)}\, ,
\ea
and  thus also 
\be
\bar {\cal M}_2 =\frac{\ln(1-\zeta_2{\bar\zeta}_1)}{(1-|\zeta_2|^2)(1-
|\zeta_1|^2)}\,.
\ee
As a result,  we finish  with 
\ba
\bra[\el_{12}]H_{12}\ket[\el_{12}]&=&\ln
\frac{(1-\zeta_1{\bar\zeta}_2)(1-\zeta_2{\bar\zeta}_1)}
{(1-|\zeta_2|^2)(1-|\zeta_1|^2)} \ , 
\ea
or, equivalently, using \rf{jh}, 
\be 
\bra[\el_{12}]H_{12}\ket[\el_{12}]
= \ln \frac{1-\eta_{ij}\el_1^i\el_2^j}{2} \ . \ee


\end{document}